\documentclass[11pt, letterpaper]{article}

\usepackage[margin=1in]{geometry}
\usepackage[utf8]{inputenc}
\usepackage[T1]{fontenc}
\usepackage{authblk}

\usepackage{amsmath,amssymb,amsfonts,amsthm}
\usepackage{mathrsfs}
\usepackage{braket}

\usepackage{graphicx}
\usepackage{multirow}
\usepackage{booktabs}
\usepackage{tabularx}
\usepackage{array}
\usepackage{threeparttable}
\usepackage{tablefootnote}
\usepackage{placeins}

\usepackage{algorithm}
\usepackage{algorithmicx}
\usepackage{algpseudocode}
\usepackage{listings}

\usepackage[title]{appendix}
\usepackage{xcolor}
\usepackage{textcomp}
\usepackage{manyfoot}
\usepackage{hyperref} 
\usepackage{orcidlink} 
\usepackage{microtype}

\begin{document}
\sloppy

\title{A Physics-Informed Neuro-Fuzzy Framework for Quantum Error Attribution}

\author[1]{Marwa R. Hassan \orcidlink{0009-0000-9050-6854}}
\author[1]{Naima Kaabouch \orcidlink{0000-0002-1333-3352}} 
\affil[1]{University of North Dakota, Grand Forks, North Dakota, USA}

\date{}

\maketitle

\begin{abstract}
As quantum processors scale beyond 100 qubits, distinguishing software bugs from stochastic hardware noise becomes a critical diagnostic challenge. We present a neuro-fuzzy framework that addresses this attribution problem by combining Adaptive Neuro-Fuzzy Inference Systems (ANFIS) with physics-grounded feature engineering. We introduce the Bhattacharyya veto, a hard constraint calibrated to the empirically characterized noise floor of the device, which classifies as buggy any output distribution whose divergence from the ideal exceeds what the device's noise produces under normal operation. The framework is validated on two IBM Heron r2 processors. On \texttt{ibm\_fez}, across 105 circuits spanning 17 algorithm families, it attains 89.5\% accuracy (Wilson 95\% CI [82.2, 94.0]) with no abstentions; every residual error is a bug whose ideal computational-basis distribution is identical to the intended circuit's, or whose divergence lies below the device's noise floor. On a held-out suite of 30 deeper circuits (ISA depth to 100) executed on \texttt{ibm\_kingston}, with all thresholds frozen, it attains 96.7\% effective and 93.3\% strict accuracy. We resolve key ambiguities---such as distinguishing correct Grover amplification from bug-induced collapse---and identify fundamental limits of single-basis diagnostics, including a Z-basis blind spot where phase-only errors remain statistically invisible. The extracted rule base is analyzed rather than asserted: the learned rules reproduce the physics of the veto, of the Grover boundary, and of depth-dependent noise without any of them being hard-coded. This work establishes an interpretable diagnostic layer that prevents error mitigation from being applied to logically flawed circuits.

\end{abstract}

\section{Introduction}
As quantum processors scale beyond the 100-qubit regime, the field is transitioning from proof-of-concept demonstrations toward early forms of practical utility. IBM’s deployment of 156-qubit Heron processors and Google’s demonstration of neural error decoding with AlphaQubit \cite{bauschLearningHighaccuracyError2024b} exemplify this transition. This scaling, however, introduces a fundamental challenge: when a quantum program fails to produce the expected outcome, determining whether the failure stems from a software bug in the circuit design or from hardware noise in the quantum processor has become increasingly difficult.

These neural decoders operate on the premise that the logical circuit itself is correct—they cannot distinguish a noisy but logically correct program from a noise-free but logically flawed one. Our framework addresses the complementary problem of error attribution: determining whether unexpected outputs arise from hardware noise (amenable to error correction) or software bugs (requiring logical debugging). This diagnostic step is a necessary precursor to effective error correction.

Quantum programs are probabilistic by nature; a failure is not a crash but a deviation 
in the probability distribution of measurement outcomes. These deviations arise from 
three physically distinct sources: stochastic hardware noise ($T_1$/$T_2$ decoherence, 
readout errors), systematic hardware errors (crosstalk, calibration drift), and software 
bugs (wrong gates, incorrect angles, missing operations) 
\cite{krantzQuantumEngineersGuide2019, sheldonCharacterizingErrorsQubit2016, huangStatisticalAssertionsValidating2019a}. The central diagnostic challenge is that systematic hardware errors and software bugs produce similar statistical signatures: both are deterministic, preserve state purity, and cause quadratic fidelity decay with circuit depth \cite{sheldonCharacterizingErrorsQubit2016}. Our framework addresses this ambiguity by classifying deviations as either HARDWARE-induced (suggesting error mitigation) or SOFTWARE-induced (requiring code debugging).

Preskill’s foundational work on Noisy Intermediate-Scale Quantum (NISQ) computing \cite{preskillQuantumComputingNISQ2018a} anticipated this challenge, noting that noise in quantum gates constrains the scale and reliability of executable quantum circuits. What was perhaps less anticipated was how this noise would complicate the fundamental task of determining whether a program is correct.

Recent work on testing and debugging quantum programs identifies the difficulty of distinguishing software faults from hardware noise as a central open challenge for NISQ‑era quantum software engineering \cite{leiteramalhoTestingDebuggingQuantum2025, murilloQuantumSoftwareEngineering2025a}. Ramalho et al. \cite{leiteramalhoTestingDebuggingQuantum2025} emphasize that, for quantum programs, defining suitable oracles and separating incorrect outputs from noise‑induced deviations is particularly challenging, and list this among their key research challenges for the field through 2030. While machine learning approaches like QOIN \cite{muqeetMitigatingNoiseQuantum2024c} have demonstrated success in filtering noise patterns, they do not address attribution—see Section \ref{sec:statistical} for detailed comparison.

\subsection{The Scale of the Problem}
The attribution problem is not merely inconvenient—it fundamentally undermines the quantum software development process. Empirical studies have documented the prevalence and complexity of bugs in quantum software: Luo et al. \cite{luoComprehensiveStudyBug2022} analyzed 96 real-world bugs across Qiskit, Cirq, Q\#, and ProjectQ, finding that over 80\% were related to quantum-specific components. Paltenghi and Pradel \cite{paltenghiBugsQuantumComputing2022b} studied 223 bugs from 18 open-source quantum computing projects, revealing that 39.9\% occurred in quantum-specific code and—critically—many do not cause crashes but silently produce erroneous results, making them difficult to detect. Zhao et al. \cite{zhaoIdentifyingBugPatterns2021a} identified eight distinct bug patterns in Qiskit programs, including incorrect gate operations, measurement issues, and improper qubit initialization.

These bugs interact with hardware noise in complex ways. Huang and Martonosi \cite{huangStatisticalAssertionsValidating2019a} introduced statistical assertions for quantum programs, categorizing bugs into algorithmic, coding, and compilation issues. Their work revealed that even expert quantum programmers frequently introduce bugs that are masked or amplified by hardware noise, making root cause analysis challenging. Aoun et al. \cite{elaounBugCharacteristicsQuantum2023a} further showed that quantum-related bugs are often more costly to fix than classical bugs, requiring more extensive code changes—underscoring the importance of early and accurate attribution.

Current debugging workflows typically involve one of three approaches, each with important limitations. First, developers may use noise-free simulation to establish ground truth, but this is computationally intractable for utility-scale circuits. Second, developers may employ quantum process tomography, but this requires exponentially many measurements. Third, developers often default to applying error mitigation blindly—an approach that, as discussed in the next subsection, can actively conceal software defects rather than reveal them.

Furthermore, the reliance on classical simulation as a `ground truth' oracle will not scale indefinitely. As demonstrated by the `quantum utility' experiments on 127-qubit processors \cite{kimEvidenceUtilityQuantum2023}, the field is entering a regime where quantum circuits yield expectation values beyond the reach of brute-force classical verification. The present work assumes a reference distribution is available, which holds for the 2--6 qubit circuits studied here and for the sub-circuits into which larger programs are routinely decomposed for testing; removing that dependence---through partial simulation, classical shadows, or scaled-down instances---is the open problem that follows from it.

Existing debugging techniques face additional scalability challenges. Metwalli and Van Meter \cite{metwalliToolDebuggingQuantum2022a, metwalliTestingDebuggingQuantum2024a} developed Cirquo, a slicing-based approach that divides quantum circuits into smaller blocks for isolated analysis. While effective for small circuits, this approach becomes increasingly complex to manage as circuit size grows. Liu et al. \cite{liuQuantumCircuitsDynamic2020a} introduced dynamic assertions using ancilla qubits to infer quantum states without direct measurement, but these can only verify specific properties (classical values, superposition, or particular entanglement patterns) rather than provide comprehensive state inspection. The fundamental constraint remains the no-cloning theorem: unlike classical debugging where variables can be freely inspected, quantum states cannot be copied or observed without disturbance \cite{miranskyyYourQuantumProgram2020}.

\subsection{The Limits of Current Error Mitigation}
State-of-the-art error mitigation techniques, such as Zero-Noise Extrapolation (ZNE) \cite{temmeErrorMitigationShortDepth2017, giurgica-tironDigitalZeroNoise2020} and Probabilistic Error Cancellation (PEC) \cite{vandenbergProbabilisticErrorCancellation2023, guptaProbabilisticErrorCancellation2024b}, focus on estimating the noise-free expectation value of observables \cite{liaoMachineLearningPractical2024c}. These methods are agnostic to the source of the error; they simply attempt to invert the noise channel. If a software bug is present, ZNE will extrapolate the incorrect logical result to the zero-noise limit, effectively mitigating the noise to reveal a perfectly wrong answer with high confidence. This highlights the critical need for error attribution—a diagnostic step that precedes mitigation—to identify whether the underlying circuit logic is sound.

Crucially, error-mitigation methods are designed to remove noise from correct circuits; they are not procedures for detecting logical bugs. As Bultrini et al. \cite{bultriniUnifyingBenchmarkingStateoftheart2023} demonstrate, applying mitigation to a buggy circuit produces high-fidelity estimates of the incorrect state. This creates a misleading appearance of correctness (high purity) that conceals the underlying programmatic error rather than revealing it.

This failure mode is the attribution gap: a diagnostic blind spot in which a developer cannot distinguish between a noisy execution of a correct circuit and a noise-mitigated execution of a faulty one. Addressing this gap is therefore not simply an optimization concern, but a necessary condition for building reliable quantum software systems.

The use of classical artificial intelligence to support quantum computing is
an established direction. The recent survey of Acampora et al.~\cite{acamporaQuantumArtificialIntelligence2026} organizes it into three
areas---quantum compilation, quantum characterization, and noise
reduction---each concerned with executing a circuit more reliably, and each
presupposing that the circuit itself is correct. Attribution, the prior
question of whether an observed deviation originates in the device or in the
program, does not appear as a category in that taxonomy. It is the gap this
work addresses.

Recent neural network approaches to quantum error handling (reviewed in Section \ref{sec: machine learning}) demonstrate the power of machine learning in this domain, but focus on error correction rather than attribution.

However, applying machine learning to error attribution presents unique challenges not faced in error correction or mitigation. The classification boundary between "hardware noise" and "software bug" is inherently fuzzy—there is no sharp threshold where one becomes the other. Some circuits are ambiguous by nature: a Grover search algorithm that correctly amplifies the target state produces a highly peaked distribution that might look like a bug to a classifier expecting uniform outputs. Conversely, certain bugs produce distributions that closely resemble what noise would create. Any practical attribution system must handle this inherent ambiguity carefully \cite{huangStatisticalAssertionsValidating2019a}.

\subsection{Proposed Approach: Physics-Informed Neuro-Fuzzy Classification}

These challenges are addressed through a proposed hybrid neuro-fuzzy framework that combines three key elements. First, an Adaptive Neuro-Fuzzy Inference System (ANFIS) \cite{jangANFISAdaptivenetworkbasedFuzzy1993} is employed that learns the classification boundary from data while maintaining interpretability through explicit fuzzy rules. Unlike black-box neural networks, ANFIS produces rules that can be inspected and understood—for example, "IF entropy deviation is HIGH and Bhattacharyya distance is HIGH THEN classify as BUG." This interpretability is crucial for building developer trust in the system's recommendations, aligning with calls for explainable AI in safety-critical quantum development \cite{leiteramalhoTestingDebuggingQuantum2025}.

Second, seven features are engineered that capture complementary aspects of quantum circuit execution. Shannon entropy and bias metrics characterize the overall shape of output distributions. The Bhattacharyya distance \cite{bhattacharyyaMeasureDivergenceTwo1943} measures topological divergence between measured and expected outputs, a choice grounded in quantum state distinguishability (Section~\ref{sec: features}).

Third, the learned classifier is augmented with a physics-informed constraint we call the Bhattacharyya veto. Hardware noise---decoherence, gate errors, and readout imperfections---was not observed to move a correctly-executed circuit's output more than a small distance from the ideal; the veto classifies as buggy any circuit whose divergence exceeds that empirically characterized floor by a wide margin, regardless of what the neural classifier predicts. The veto is sound but not complete: crossing it implies a fault, but a fault need not cross it.

\subsection{Contributions}
This paper makes the following contributions:
\begin{enumerate}
    \item Formalization of the Attribution Gap: We characterize error attribution as a distinct diagnostic problem separate from mitigation or correction, and characterize the bounds on distinguishing coherent logic errors from stochastic noise given a reference output distribution.
    \item The Bhattacharyya Veto: We introduce a hard constraint calibrated to the empirically characterized noise floor of the device. Across 95 correctly-executed circuits on two Heron r2 processors, spanning ISA depths from 1 to 101, the largest value observed is $D_B^{\mathrm{log}} = 0.057$, apart from a single circuit at $0.246$ recorded while the backend reported a maintenance status, and the floor is flat with depth. The veto threshold sits $3.5\times$ above that value and produced one false positive, that same circuit, in 95 correct circuits.
    \item Resolution of the Grover Classification Boundary: We identify and resolve a failure mode where correct amplitude amplification mimics the statistical signature of coherent faults, and show that the trained rule base contains an explicit ``peaked-by-design'' rule that encodes the resolution.
    \item Hardware Validation on Two Current-Generation Devices with a Held-Out Test: We validate on \texttt{ibm\_fez} across 105 circuits spanning 17 algorithm families (89.5\% accuracy), and on a held-out suite of 30 deeper circuits on \texttt{ibm\_kingston}, with every threshold frozen before the data were seen (96.7\% effective, 93.3\% strict). To our knowledge, this is one of the first studies of error attribution on production quantum hardware to report a held-out, second-device result.
    \item \textbf{An Audited Training Pipeline:} We identify three properties a 
    synthetic training set for attribution must have---computing structural features 
    on the transpiled circuit, simulating every sample under device noise, and 
    filtering out injected bugs whose ideal output is indistinguishable from the 
    correct circuit---and show what fails without them (Section~\ref{sec: corrections}). The third is substantial: 40\% of mutation-style bug injections produce no computational-basis change, and would otherwise be labeled as bugs.
\end{enumerate}

The remainder of this paper is organized as follows. Section 2 provides background on quantum errors and related work in machine learning for quantum computing. Section 3 details our methodology, including feature extraction, ANFIS architecture, and the physics-informed veto mechanism. Section 4 describes the experimental setup on IBM Quantum hardware. Section 5 presents results and analysis. Section 6 discusses limitations and the physics boundaries encountered. Section 7 concludes with future directions.

\section{Background And Related Work}
This section establishes the technical foundations for the proposed approach. The discussion begins with the physics of quantum errors, distinguishing incoherent noise from coherent errors. Relevant machine learning approaches for quantum computing are then reviewed, followed by background on neuro-fuzzy systems and classification with rejection.
\subsection{Quantum Error Taxonomy}

Errors in quantum computing arise from three physically distinct sources: 
incoherent noise from environmental decoherence, coherent errors from systematic 
control imperfections, and software bugs from mistakes in circuit construction. 
Understanding these distinctions is crucial because each error type manifests 
differently in measurement statistics and requires different remediation strategies. 
The first two sources are hardware-related and addressed through error mitigation 
or improved calibration; the third requires debugging the circuit logic itself.

\subsubsection{Incoherent Noise}
Incoherent noise arises from uncontrolled interactions between qubits and their environment, causing irreversible loss of quantum information. The dominant mechanisms in superconducting transmon qubits are $T_1$ relaxation and $T_2$ dephasing.
\begin{itemize}
    \item $T_1$ relaxation (amplitude damping) describes the decay of excited states to the ground state. This process is physically caused by spontaneous emission, two-level system (TLS) defects in the qubit substrate, and quasiparticle tunneling \cite{krantzQuantumEngineersGuide2019, chenPhononEngineeringAtomicscale}. The amplitude damping channel transforms the density matrix $\rho$ as:
     $$\rho \rightarrow E_0 \rho E_0^\dagger + E_1 \rho E_1^\dagger$$
    where $E_0 = |0\rangle\langle0| + \sqrt{1-\gamma}|1\rangle\langle1|$ and $E_1 = \sqrt{\gamma}|0\rangle\langle1|$, with $\gamma = 1 - e^{-t/T_1}$ \cite{krantzQuantumEngineersGuide2019, ahmedComparativeAnalysisNoise2025a}. This process increases entropy by mixing the state, a signature our framework detects via the Entropy Deviation feature.
    \item $T_2$ dephasing encompasses pure dephasing (characterized by $T_{\phi}$) arising from $1/f$ flux noise and coupling to a bath of TLS defects. The relationship between these timescales is $1/T_2 = 1/(2T_1) + 1/T_{\phi}$. Burnett et al. \cite{burnettDecoherenceBenchmarkingSuperconducting2019} and Carroll et al. \cite{carrollDynamicsSuperconductingQubit2022} have extensively characterized these noise sources in superconducting systems, demonstrating that fluctuations in $T_1$ can vary by factors of 2--3 over timescales of hours.
    \item The depolarizing channel provides a phenomenological model that uniformly contracts the Bloch sphere \cite{chengSimulatingNoisyQuantum2021}. For a single-qubit gate with error rate $\epsilon$, the depolarizing channel acts as:
    $$\rho \rightarrow (1-\epsilon)\rho + \frac{\epsilon}{3}(X\rho X + Y\rho Y + Z\rho Z)$$
\end{itemize}

A critical observation from Sheldon et al. \cite{sheldonCharacterizingErrorsQubit2016} is that incoherent noise causes fidelity to decay linearly with circuit depth, creating diffusive "flattening" of output distributions. This linear decay is a signature that distinguishes stochastic noise from systematic errors.

\subsubsection{Coherent Errors}
Coherent errors represent systematic unitary deviations from ideal gate operations. Unlike incoherent noise, these errors are deterministic and repeatable. Common sources include miscalibrated pulse amplitudes, frequency detuning, and cross-talk \cite{iversonCoherenceLogicalQuantum2020}. Blume-Kohout et al. \cite{blume-kohoutTaxonomySmallMarkovian2022} developed a comprehensive taxonomy distinguishing Hamiltonian errors (unitary deviations), stochastic Pauli errors, and non-Pauli stochastic errors.

Cross-talk presents a particularly challenging form of coherent error. Rudinger et al. \cite{rudingerExperimentalCharacterizationCrosstalk2021c} used simultaneous gate set tomography to characterize context-dependent crosstalk errors, demonstrating that operations on one qubit can influence error behavior on neighboring qubits. IBM’s Heron processors employ tunable couplers in a heavy-hex architecture, a design choice aimed at suppressing crosstalk and improving two-qubit gate performance relative to previous fixed-coupling systems \cite{abughanemIBMQuantumComputers2025} . Sheldon et al. \cite{sheldonCharacterizingErrorsQubit2016} further established that coherent errors cause fidelity to decay quadratically with circuit depth, as errors accumulate coherently rather than averaging out.

\subsubsection{Software Bugs in Quantum Circuits}
Software bugs differ fundamentally from hardware errors in that they represent mistakes in circuit construction rather than imperfect execution of correct circuits. Huang and Martonosi \cite{huangStatisticalAssertionsValidating2019a} categorized quantum software bugs into three classes: algorithmic bugs (incorrect oracle construction, wrong iteration counts), coding bugs (missing gates, wrong angles, wrong targets), and compilation bugs (errors introduced during transpilation).
Mutation testing tools such as Muskit \cite{mendiluzeMuskitMutationAnalysis2022} and QMutPy \cite{fortunatoQMutPyMutationTesting2022a} have revealed that many common bugs produce output distributions difficult to distinguish from hardware noise without careful statistical analysis.
Critically, both coherent hardware errors and software bugs share the property of being unitary transformations that preserve state purity \cite{giananiDiagnosticsQuantumgateCoherences2023}. Both cause fidelity to decay quadratically with depth and can shift probability mass to incorrect states—producing structured deviations—without the diffusive flattening characteristic of incoherent noise \cite{samosFidelityDecayError2025}. This distinct scaling law (quadratic vs. linear decay) is what allows our framework to distinguish coherent errors from incoherent noise. The key distinction between hardware coherent errors and software bugs is magnitude: hardware coherent errors are typically small-angle rotations $(\epsilon\ll1 rad)$ constrained by calibration tolerances \cite{lazarCalibrationDriveNonlinearity2023}, while software bugs often manifest as large-angle rotations $(\theta=\pi/2,\pi)$ that dramatically alter output topology \cite{zindorfEfficientImplementationMulticontrolled2025}. Our framework exploits this magnitude difference through the Bhattacharyya veto, which identifies topological shifts exceeding the noise floor, while acknowledging that small-angle software bugs may be indistinguishable from coherent hardware noise.

\subsection{Machine Learning for Quantum Computing} \label{sec: machine learning}
Machine learning has emerged as a powerful tool for various quantum computing tasks, from error correction to state tomography. We review the most relevant approaches in neural decoding, error mitigation, and statistical bug detection to contextualize the proposed framework.

\subsubsection{Neural Network Error Decoders}
Google's AlphaQubit \cite{bauschLearningHighaccuracyError2024b} represents the current state-of-the-art in neural network quantum error correction. The system employs a recurrent transformer architecture that processes "soft" I/Q readout signals rather than hard-threshold detection events. On Google's Sycamore processor, AlphaQubit achieves 6\% fewer errors than tensor network methods and 30\% fewer than correlated matching. The two-stage training approach—pretraining on billions of synthetic samples followed by fine-tuning on limited experimental data—demonstrates that neural networks can generalize to real hardware behavior despite training primarily on simulation.

Earlier work by Torlai et al. \cite{torlaiNeuralnetworkQuantumState2018} pioneered the use of restricted Boltzmann machines for neural network quantum state tomography, demonstrating efficient reconstruction of highly entangled states with over 100 qubits. Ahmed et al. \cite{ahmedQuantumStateTomography2021} extended this approach using conditional generative adversarial networks (GANs), achieving orders of magnitude speedup over maximum likelihood estimation. Schmale et al. \cite{schmaleEfficientQuantumState2022} showed that convolutional neural networks can reduce estimation error by an order of magnitude compared to linear inversion methods.

\subsubsection{Machine Learning for Error Mitigation}
Zero-noise extrapolation (ZNE), introduced by Temme et al. \cite{temmeErrorMitigationShortDepth2017} , artificially amplifies noise and extrapolates to the zero-noise limit. Liao et al. \cite{liaoMachineLearningPractical2024c} demonstrated that random forest models can learn the noise-scaling relationship, reducing ZNE runtime overhead by over 2× while maintaining accuracy on IBM systems up to 100 qubits. Their approach uses circuit features (depth, gate counts, qubit connectivity) to predict optimal noise scaling factors.

Probabilistic error cancellation (PEC), also from Temme et al. \cite{temmeErrorMitigationShortDepth2017}, represents noisy gates as quasi-probability distributions over ideal gates. Van den Berg et al. \cite{vandenbergProbabilisticErrorCancellation2023} developed sparse Pauli-Lindblad models that make PEC practical for larger circuits by exploiting the locality of noise. These mitigation techniques are complementary to our attribution approach—once a circuit is confirmed bug-free, mitigation can be applied with confidence.

Fuzzy methods have been applied to quantum error problems before, though to a
different task. Acampora and Vitiello~\cite{acamporaErrorMitigationQuantum2021}
were the first to use a fuzzy method---fuzzy C-means clustering---to identify
measurement-error mitigation matrices, reporting stronger error reduction than
the matrices conventionally used in Qiskit. In related work, the same group
learned the mitigation matrix with a genetic algorithm whose fitness function
is the Bhattacharyya distance~\cite{acamporaGeneticAlgorithmsBased2021}, and
the clustering approach was subsequently given a proof-of-principle validation
on a two-qubit register of a five-qubit transmon processor by Ahmad et
al.~\cite{ahmadMitigatingErrorsSuperconducting2024}. The survey of Acampora et
al.~\cite{acamporaQuantumArtificialIntelligence2026} classifies this line of
work under error mitigation.

These methods \emph{mitigate} a known error channel, and where they use $D_B$
it serves as an optimization objective. The present work \emph{attributes} an
observed deviation to its source, and uses $D_B$ as a discriminative feature.
Attribution is the prior question: a mitigation strategy cannot be selected for
a fault that originates in the program rather than the device.

\subsubsection{Statistical Approaches to Bug Detection}\label{sec:statistical}
The application of machine learning to quantum software quality has expanded rapidly, though the specific challenge of error attribution remains distinct from noise mitigation. Muqeet et al. \cite{muqeetMitigatingNoiseQuantum2024c, pontolilloIdealNoisyAdapting2025} developed QOIN, which trains neural networks to learn noise patterns specific to quantum computers and individual circuits, then applies transfer learning to filter noise from test outputs. On evaluation across IBM's 23 noise models, Google's two available noise models \cite{aruteQuantumSupremacyUsing2019a}, and Rigetti's Quantum Virtual Machine \cite{smithPracticalQuantumInstruction2017}, QOIN reduced the noise effect by more than 80\% on most models. However, its goal—noise removal—differs from attribution. QOIN produces filtered distributions that are passed to an existing test oracle for pass/fail assessment; it does not decide whether residual deviation is due to a software bug or an unlearned noise pattern.
Alternative approaches avoid machine learning entirely: AssertsQ \cite{veganzonesAssertsQQuantumAssertion2025a} inserts runtime checks (e.g., swap tests) directly into circuits, but this adds gate overhead and potential noise sources.

Virani et al. \cite{viraniDistinguishingQuantumSoftware2025a} proposed the Bias-Entropy Model, a statistical approach that distinguishes bugs from noise by comparing Most Probable States (MPS) against Desired States (DS). Their key insight is that hardware noise increases entropy while preserving MPS below a threshold, whereas software bugs shift MPS away from DS. However, their validation was limited to quantum simulators with shallow circuits ($\leq$15 qubits, depth 1--5), and their method relies on fixed statistical thresholds that—as the authors acknowledge—struggle when ``the entropy deviation from bugs is highly dependent on specific bug and circuit characteristics.'' 

Our framework extends this foundation in three ways: (1) we validate on production quantum hardware (two IBM Heron r2 processors) rather than simulators; (2) we replace rigid thresholds with adaptive fuzzy membership functions learned from data, incorporating seven physics-grounded features including entropy \emph{deviation} (divergence from expected rather than absolute entropy) and Bhattacharyya distance; and (3) we introduce the Bhattacharyya veto, a hard constraint calibrated to the empirically characterized noise floor of the device, which statistical threshold approaches lack. These enhancements enable resolution of edge cases---such as the Grover boundary---where bias-entropy thresholds alone fail.

\subsection{Adaptive Neuro-Fuzzy Inference Systems}
The Adaptive Neuro-Fuzzy Inference System (ANFIS), introduced by Jang \cite{jangANFISAdaptivenetworkbasedFuzzy1993}, combines the learning capability of neural networks with the interpretability of fuzzy logic systems. With over 16,900 citations, ANFIS remains one of the most influential architectures for problems requiring both accuracy and explainability.

ANFIS implements a first-order Takagi-Sugeno fuzzy inference system \cite{takagiFuzzyIdentificationSystems1985} through a five-layer neural network. The architecture consists of: (1) a fuzzification layer with parameterized Gaussian membership functions, (2) a rule firing layer computing T-norm products of membership degrees, (3) a normalization layer, (4) a defuzzification layer with linear consequent functions, and (5) an aggregation layer producing the final output.

A key advantage of ANFIS over black-box neural networks is rule extraction. The trained network can be interpreted as a set of fuzzy if-then rules, such as ``IF \texttt{entropy\_deviation} is HIGH AND \texttt{bhattacharyya\_distance} is HIGH THEN output is BUG.'' This interpretability is crucial for building developer trust. Developers can inspect why the system made a particular attribution rather than accepting an opaque prediction.

\subsection{Classification with Rejection}
Classification with rejection—also called selective classification \cite{chowOptimumRecognitionError1970} or classification with abstention—allows a classifier to decline making predictions when confidence is insufficient. This capability is essential for error attribution, where overconfident incorrect predictions could lead developers to waste time debugging phantom bugs or ignoring real ones.

The theoretical foundations were established by Chow \cite{chowOptimumRecognitionError1970}, who derived the Bayes-optimal error-reject tradeoff. Bartlett and Wegkamp \cite{bartlettClassificationRejectOption2008} introduced the double hinge loss as a convex surrogate enabling SVM-style learning with rejection. For deep learning, Geifman and El-Yaniv \cite{geifmanSelectiveClassificationDeep2017, geifmanSelectiveNetDeepNeural2019} developed SelectiveNet, an architecture that jointly learns classification and rejection, achieving state-of-the-art risk-coverage tradeoffs—for example, 2\% top-5 error on ImageNet with 60\% coverage.

Beyond statistical optimality, the three-class design reflects broader principles from safety-critical systems engineering. Leveson \cite{levesonEngineeringSaferWorld2016} argues that in complex systems where failures carry significant consequences, explicitly modeling uncertainty is preferable to forcing binary decisions that may be confidently wrong. In quantum software debugging, a misclassification in either direction—labeling a bug as noise, or noise as a bug—wastes developer resources and may mask genuine faults. The uncertain class operationalizes this principle: it signals "insufficient evidence for confident attribution" rather than forcing a potentially harmful guess.

The proposed approach draws on this literature by defining an explicit "uncertain" class for samples where the classifier's confidence is insufficient. We employ asymmetric thresholds that reflect the different costs of misclassifying bugs versus noise, following the Bayesian risk minimization framework. Hand and Till \cite{handSimpleGeneralisationArea} provided methods for extending ROC analysis to multi-class problems, which we adapt for evaluating our three-class (bug/noise/uncertain) classifier.

\subsection{Statistical Measures for Distribution Comparison}
Our feature engineering draws on statistical measures for comparing probability distributions. Shannon entropy \cite{shannonMathematicalTheoryCommunication1948} quantifies uncertainty: $H(X) = -\sum_{i} p_i \log p_i$. For quantum systems, the von Neumann entropy $S(\rho) = -\mathrm{Tr}(\rho \ln \rho)$ extends this to density matrices. Importantly, quantum entropy exhibits non-monotonicity: the entropy of a composite system can be zero while subsystem entropies are positive, a signature of entanglement.

The Bhattacharyya coefficient $BC(P,Q) = \sum_{x} \sqrt{P(x)Q(x)}$ measures distributional overlap, with the Bhattacharyya distance defined as $D_B = -\ln(BC)$ \cite{bhattacharyyaMeasureDivergenceTwo1943}. This metric has deep connections to quantum state distinguishability: the quantum fidelity equals the square of the minimum Bhattacharyya coefficient over all measurements \cite{fuchsCryptographicDistinguishabilityMeasures1999}. The related Hellinger distance $H^2(P,Q) = 1 - BC(P,Q)$ provides a true metric satisfying the triangle inequality.

These measures connect to the fundamental limits of quantum state discrimination. Helstrom \cite{helstromQuantumDetectionEstimation1969} established that the minimum error probability for distinguishing two quantum states is $P_{\mathrm{err}} = \frac{1}{2}(1 - D(\rho,\sigma))$, where $D$ is the trace distance. Our use of Bhattacharyya distance is thus grounded in quantum information theory---when this distance is large, the measured distribution is distinguishable from what noise alone could produce.

\section{METHODOLOGY}
The proposed framework consists of three integrated components: (1) a physics-grounded feature extraction pipeline that transforms raw circuit execution results into discriminative metrics, (2) an ANFIS classifier that learns the attribution boundary from labeled examples, and, (3) a physics-informed veto mechanism that enforces physical constraints on the classifier's outputs. 

\subsection{Physics-Grounded Feature Engineering}\label{sec: features}
Effective error attribution requires features that capture the different ways hardware noise and software bugs manifest in output distributions. Rather than feeding raw bitstrings or amplitude vectors into the model, the framework extracts high-level thermodynamic and information-theoretic metrics. These serve as proxies for underlying density matrix properties (purity, fidelity, coherence), enabling the classifier to discriminate between the diffusive nature of noise and the structured distortion of bugs. Seven features are extracted from each circuit execution, summarized in Table~\ref{tab:features}:
\begin{table}[ht]
  \caption{Summary of Extracted Features}
  \label{tab:features}
  \centering
  \begin{tabular}{lll}
    \toprule
    \textbf{Feature} & \textbf{Description} & \textbf{Discriminative Role} \\
    \midrule
    Entropy & Shannon entropy of output & Noise increases; some bugs decrease \\
    Bias & $L_2$ distance from uniform & Distinguishes peaked vs.\ spread outputs \\
    Max Prob & Maximum state probability & Detects loss of determinism \\
    Norm Depth & $\ln(1+depth)$ & Contextualizes noise expectations \\
    2Q Density & Two-qubit gate ratio & Indicates noise susceptibility \\
    Entropy Dev & $|H_{\text{measured}} - H_{\text{ideal}}|$ & Solves Grover classification problem \\
    $D_B^{\log}$ & Log Bhattacharyya distance & Topological divergence detection \\
    \bottomrule
  \end{tabular}
\end{table}
\subsubsection{Feature 1: Shannon Entropy}
Shannon entropy is the standard scalar measure of uncertainty in a discrete probability distribution and is routinely used to summarize outcome uncertainty from quantum measurement histograms \cite{guoNoiseEffectsPurity2023}.  Given measurement counts converted to a probability distribution $$p = (p_1, \dots, p_{2^n})$$
over n-qubit computational basis states, the Shannon entropy is:
$$H(p) = -\sum_{i} p_i \log_2(p_i)$$

For an n-qubit system, entropy ranges from 0 (deterministic output) to n bits (uniform distribution). Hardware noise typically increases entropy by driving the system toward a maximally mixed state through decoherence, spreading probability mass across states \cite{perezErrorDivisibleTwoQubitGates2023}. In contrast, unitary software bugs preserve state purity—a buggy circuit may produce a completely wrong output, but that output can still be highly peaked (low entropy) if the bug is a coherent rotation error. However, certain bugs (e.g., missing entangling gates) can either increase or decrease entropy depending on the circuit \cite{eisertEntanglingPowerQuantum2021, everedHighfidelityParallelEntangling2023}.

\subsubsection{Feature 2: Bias} \label{sec: bias}
Bias measures the $L_2$ distance between the measured probability distribution and the uniform distribution \cite{bussandriQuantumDistanceMeasures2023, riofrioCharacterizationQuantumGenerative2024}:
$$\beta(p)=\parallel p-u\parallel_2=\sqrt{\sum_{i=1}^{2^n}{(p_i-\frac{1}{2^n})}^2}$$
where $u=(1/2^n,\ldots,1/2^n)$ represents the maximally mixed state. This metric quantifies distributional peakedness—how concentrated probability mass is versus spread uniformly \cite{liuObservationEntanglementTransition2023}. The metric ranges from 0 (perfectly uniform) to approximately 1 (deterministic output) \cite{bussandriQuantumDistanceMeasures2023}.

Hardware noise systematically flattens distributions and reduces bias \cite{chernyavskiyEntropicPropertyRandomized2023}. Software bugs exhibit variable effects: collapse bugs (e.g., missing Hadamard) increase bias by producing unintended deterministic outputs, while scrambling bugs may decrease bias by disrupting interference patterns \cite{chernyavskiyEntropicPropertyRandomized2023}. This asymmetric response makes bias particularly valuable for resolving the Grover classification boundary, where both correct Grover circuits and collapsed buggy circuits produce high-bias outputs. Combined with entropy deviation (Feature 6: \ref{sec: feature 6}), the framework distinguishes these cases—correct circuits show high bias with low entropy deviation, while collapsed circuits show high bias with high entropy deviation.

\subsubsection{Feature 3: Maximum Probability}
Maximum probability $p_{max} = \max(p_i)$ indicates whether a single outcome dominates the distribution. Correctly-functioning deterministic algorithms should have $p_{max} \approx 1$ for some target state, while uniform circuits should have $p_{max} \approx 1/2^n$. Noise drives outputs toward this uniform limit \cite{boixoCharacterizingQuantumSupremacy2018b}, while certain bugs can create unintended structure that increases $p_{max}$.

\subsubsection{Feature 4: Normalized Circuit Depth}

Circuit depth directly correlates with noise accumulation, but this relationship is not strictly linear. The depth was encoded using logarithmic normalization:
$$ d_{\text{norm}} = \ln(1+d) $$

This transformation reflects three considerations. First, under dynamical decoupling sequences employed at optimization level 3 on IBM hardware, effective error accumulation follows sub-linear scaling rather than linear growth as circuits approach the mixed-state limit \cite{cooteResourceEfficientContextAwareDynamical2025}. Second, the logarithmic form captures relative sensitivity: a depth increase from 5 to 10 is physically more significant than an increase from 100 to 105, consistent with the diminishing returns of additional gates in noise-saturated regimes \cite{wangProspectUsingGrovers2020}. Third, this normalization prevents deep circuits from dominating the feature space, ensuring that subtle distributional features remain impactful for classification.

\subsubsection{Feature 5: Two-Qubit Gate Density}
Two-qubit gates constitute the dominant noise source in superconducting processors, with error rates approximately one order of magnitude higher than single-qubit operations \cite{liErrorSinglequbitGate2023, dingHighFidelityFrequencyFlexibleTwoQubit2023}. We compute the ratio:
$$\rho_{2Q}=\frac{N_{2Q}}{N_{total}}$$

This metric, adopted by benchmark suites such as SupermarQ \cite{tomeshSupermarQScalableQuantum2022}, reflects circuit noise susceptibility. Higher $\rho_{2Q}$ increases the prior probability that observed deviations stem from hardware noise rather than software bugs, informing the classifier's attribution confidence. Both structural features (depth and two-qubit gate density) are computed on the post-transpilation circuit, so that gate counts and depth reflect the native operations actually executed on hardware rather than the abstract pre-transpilation form.

\subsubsection{Feature 6: Entropy Deviation}\label{sec: feature 6}
This feature captures the difference between measured and expected entropy:
$$\Delta H=\mid H_{measured}-H_{expected}\mid$$

Expected entropy is computed via ideal simulation. This metric addresses the ambiguity of high-entropy algorithms: a Quantum Fourier Transform produces legitimately high entropy, so raw entropy alone would be misleading. By comparing to the expected value, deviations are isolated from algorithmic intent rather than absolute entropy values.

This feature is essential for resolving the Grover classification boundary. A correctly-implemented Grover search produces a low-entropy (peaked) distribution, as does a bug that collapses a GHZ state to a product state. However, Grover's expected entropy is low, so $\Delta H$ remains small. The collapsed GHZ state has higher expected entropy (from the intended superposition), so $\Delta H$ is large. By comparing against algorithm-specific expectations, the framework distinguishes legitimate peaked distributions from buggy ones. Combined with bias (Feature 2: \ref{sec: bias}), entropy deviation enables further disambiguation: correct Grover circuits show low entropy deviation with high bias (due to concentration), while collapsed circuits show high entropy deviation despite similar raw entropy values.

Our selection of Entropy Deviation is grounded in the thermodynamic distinguishability of unitary errors versus decoherence. Hardware noise in superconducting processors is dominated by $T_1$ relaxation and $T_2$ dephasing. While $T_1$ ultimately drives qubits toward the ground state, in the NISQ regime these processes drive the system toward higher-entropy mixed states, contrasting with the pure states preserved by unitary evolution \cite{nielsenQuantumComputationQuantum2010, krantzQuantumEngineersGuide2019}. Software bugs—such as incorrect rotation angles or wrong target qubits—manifest as unitary transformations. Being unitary, these bugs preserve state purity; they rotate the state vector to an incorrect sector of the Hilbert space without introducing decoherence \cite{nielsenQuantumComputationQuantum2010}. By monitoring the deviation $|\Delta H|$ rather than absolute entropy, the framework effectively asks a thermodynamic question: ``Is the observed error consistent with irreversible information loss (noise), or does it exhibit the purity-preserving signature of a coherent logic error?''

\subsubsection{Feature 7: Log-Transformed Bhattacharyya Distance}

The Bhattacharyya coefficient measures distributional overlap between measured and expected distributions \cite{bhattacharyyaMeasureDivergenceTwo1943}:
\[
    BC(p,q) = \sum_x \sqrt{p(x) \cdot q(x)}
\]
The Bhattacharyya distance $D_B = -\ln(BC)$ connects to quantum state distinguishability through the Fuchs-van de Graaf inequalities \cite{fuchsCryptographicDistinguishabilityMeasures1999}: the quantum fidelity satisfies $F(\rho,\sigma) \leq BC^2$, so high $D_B$ guarantees divergence from the intended state.

In our hardware validation (Section \ref{sec: results}), raw Bhattacharyya distances exhibit a heavy-tailed distribution: noise-affected circuits yield $D_B \in [0.0001, 0.043]$, while buggy circuits span $D_B \in [0.0001, 10.0]$---a range exceeding two orders of magnitude. This dynamic range destabilizes gradient descent for Gaussian membership functions. We apply logarithmic compression:
\[
    D_B^{\mathrm{log}} = \ln(1 + D_B)
\]
which maps the range $[0, 10]$ to approximately $[0, 2.4]$, preserving sensitivity near zero where noise-level discrimination occurs while compressing the heavy tail. After transformation, noise circuits fall within $D_B^{\mathrm{log}} \in [0.0001, 0.042]$ (mean 0.011) and bug circuits span $D_B^{\mathrm{log}} \in [0.0001, 2.40]$ (mean 0.415). Values exceeding $D_B^{\mathrm{log}} > 0.20$ provide the basis for the Bhattacharyya veto (Section \ref{sec: veto}).

The Bhattacharyya distance has been used in quantum error work before, as the
fitness function of a genetic algorithm learning measurement-mitigation
matrices~\cite{acamporaGeneticAlgorithmsBased2021}; here it serves instead as
a discriminative input feature.

\subsection{ANFIS Architecture: The 5-Layer Neuro-Fuzzy Stack}
The Adaptive Neuro-Fuzzy Inference System \cite{jangANFISAdaptivenetworkbasedFuzzy1993} constructs a hybrid learning architecture that combines the backpropagation capabilities of neural networks with the semantic interpretability of fuzzy logic. Unlike black-box deep neural networks that learn opaque feature representations, ANFIS utilizes a rule-based architecture that allows extraction of interpretable heuristics.

\begin{figure}[h]
     \centering
     \includegraphics[width=0.95\linewidth]{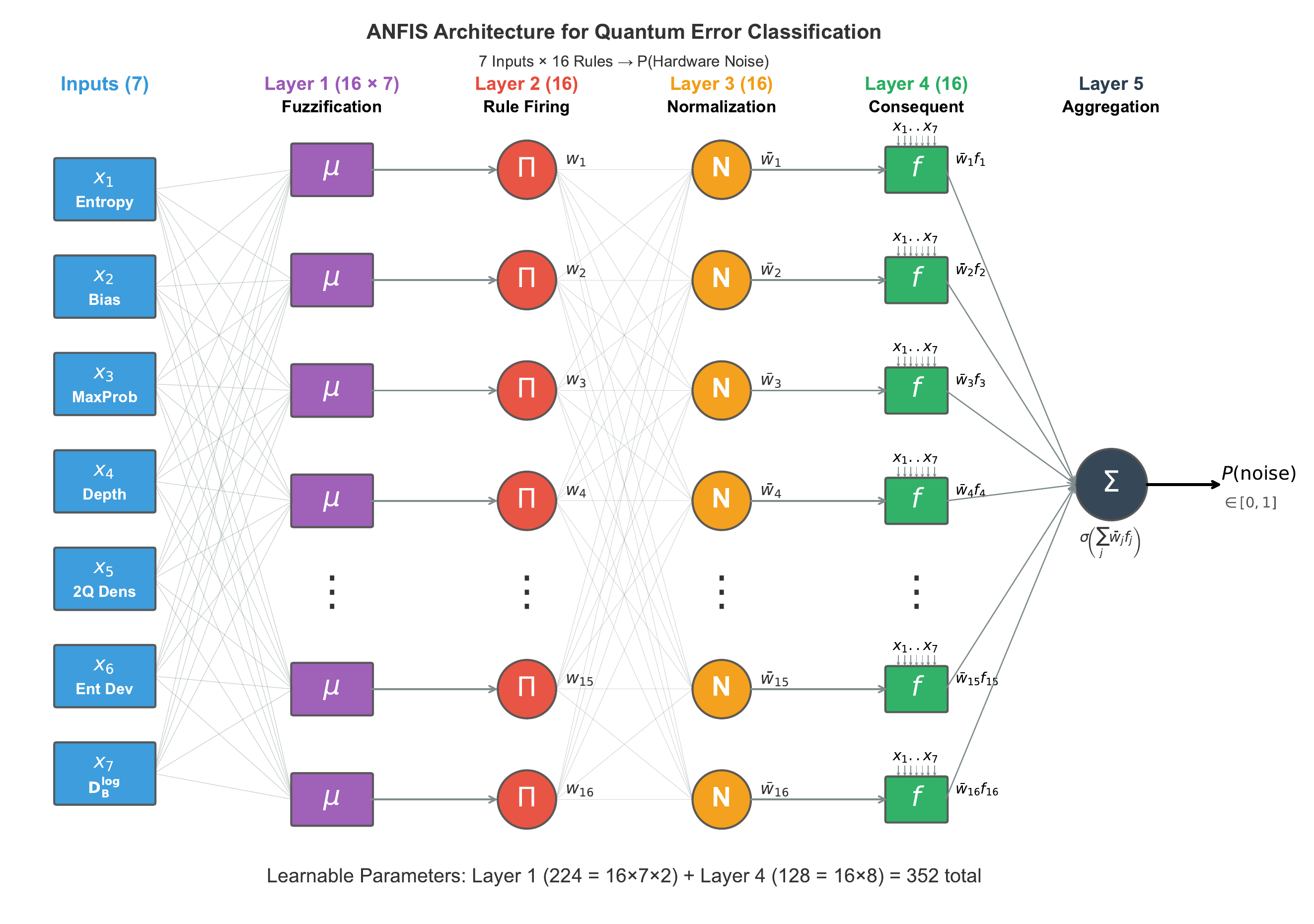}
     \caption{ANFIS Architecture: The 5-Layer Neuro-Fuzzy Stack. Each $\mu$ block bundles the seven Gaussian membership functions of one rule, so Layer 1 comprises $16 \times 7 = 112$ membership functions with $224$ learnable parameters.}
     \label{Architecture}
 \end{figure}

The proposed classifier implements a first-order Takagi-Sugeno ANFIS with 16 fuzzy rules operating on the 7 input features discussed in Section \ref{sec: features}. The architecture follows Jang's original formulation \cite{jangANFISAdaptivenetworkbasedFuzzy1993} with adaptations for binary classification with rejection. Figure \ref{Architecture}  illustrates the architecture.
The complete ANFIS model contains approximately 350 learnable parameters---a compactness that reflects the framework's reliance on physics-informed feature engineering and enables extraction of human-interpretable rules. The network structure is as follows:
\paragraph{Layer 1: Fuzzification (Semantic Layer)}
The first layer maps the 7 input features to fuzzy membership degrees using Gaussian membership functions:
\[
    \mu_{ij}(x) = \exp\left(-\frac{(x_i - c_{ij})^2}{2\sigma_{ij}^2}\right)
\]
Where $c_{ij}$ and $\sigma_{ij}$ are learnable parameters representing the centers and widths of fuzzy sets (e.g., `Low Entropy', `High Bias'). The implementation includes a data-driven initialization routine that spreads centers across the normalized feature space, avoiding local minima where all rules collapse to the same semantic meaning.
With 16 rules and 7 features, this layer contains 224 learnable parameters ($16 \times 7 \times 2$).

\paragraph{Layer 2: Rule Firing (Logical AND)}
This layer computes the firing strength $w_i$ of each rule using the product T-norm:
\[
    w_i = \prod \mu_{ij}(x_j)
\]
This mathematical operation corresponds to the logical AND; a rule fires strongly only if all its antecedent conditions are met (e.g., `IF Entropy is Low AND Bias is High AND Bhattacharyya is High'). This structure captures non-linear interactions between physics metrics: high bias alone might be ambiguous, but high bias combined with low entropy is a strong signature of a coherent bug.

\paragraph{Layer 3: Normalization}
This layer normalizes firing strengths: $\bar{w}_i = w_i / \sum w_k$, ensuring that rules compete for influence. This effectively partitions the feature space into soft regions governed by different rules.

\paragraph{Layer 4: Takagi-Sugeno-Kang (TSK) Consequent}
Unlike Mamdani systems that output fuzzy sets \cite{mamdaniExperimentLinguisticSynthesis1975}, the TSK model \cite{takagiFuzzyIdentificationSystems1985} outputs a linear function of the inputs:
\[
    f_i = p_{i0} + \sum p_{ij}x_j
\]
Each rule governs a local region of the physics feature space and applies a specific linear classifier within that region. This local linearity is more powerful than global logistic regression, enabling the model to fit complex, non-linear decision boundaries in quantum error landscapes.
\paragraph{Layer 5: Aggregation}
The final output is a weighted sum passed through a sigmoid: $y = \sigma(\sum \bar{w}_i f_i)$, providing a probability score $P(\text{Noise})$. Values near 0 indicate high confidence of a software bug; values near 1 indicate high confidence of hardware noise; values near 0.5 are flagged as uncertain.

\subsection{Uncertainty-Penalized Training}
\paragraph{Initialization}
Membership function centers are initialized by randomly sampling 16 training examples, spreading the rules across the feature space \cite{jangANFISAdaptivenetworkbasedFuzzy1993}. Widths are set proportional to each feature's standard deviation, scaled by a random factor in $[0.5, 1.0]$ to encourage rule diversity. Consequent parameters are initialized with small random values ($\sigma=0.1$).

\paragraph{Loss Function}
An uncertainty-penalized loss is employed that combines weighted binary cross-entropy with an explicit penalty for indecisive predictions:
\[
    \mathcal{L} = \mathcal{L}_{\mathrm{BCE}}^w + \lambda \cdot \max(0, \min(p, 1-p) - \tau_{\mathrm{margin}})
\]

The first term is class-weighted Binary Cross-Entropy (BCE):
\[
    \mathcal{L}_{\mathrm{BCE}}^w = -\sum_i w_i [y_i \ln(p_i) + (1-y_i) \ln(1-p_i)]
\]
where $w_i = N_{\mathrm{total}}/(2N_{c_i})$ assigns higher weight to the minority class, handling imbalance between bug and noise samples. The second term quantifies prediction uncertainty as $\min(p, 1-p)$, which peaks at 0.5 (maximum ambiguity) and vanishes at 0 or 1 (full confidence). With $\tau_{\mathrm{margin}}=0.20$ and $\lambda=0.3$, the penalty activates when predictions fall within the uncertain zone ($0.20 < p < 0.80$), creating a repulsive potential that pushes the optimization toward decisive classifications.

This formulation draws on principles from classification with rejection \cite{bartlettClassificationRejectOption2008}, adapted for the quantum verification context where silent low-confidence failures are more costly than explicit uncertainty flags.

\paragraph{Optimization}
Training uses the Adam optimizer \cite{kingmaAdamMethodStochastic2017} with learning rate 0.005 and weight decay $10^{-4}$. The model trains for up to 500 epochs with early stopping (patience = 60 epochs) based on validation accuracy. A learning rate scheduler reduces the rate by half after 20 epochs of stagnation. Full-batch gradient descent is used given the moderate dataset size (1,983 training samples).

\paragraph{Feature Normalization}
All features are standardized to zero mean and unit variance before training. The normalization parameters (mean and standard deviation for each feature) are saved with the model checkpoint for consistent inference.

\paragraph{Classification Thresholds}
Classification thresholds $\tau_{\mathrm{bug}}=0.35$ and $\tau_{\mathrm{noise}}=0.70$ were selected on the \texttt{ibm\_fez} validation suite during development and were \emph{not} re-tuned for the model reported here. They therefore function as fixed hyperparameters: every result in this paper, including the held-out evaluation of Section~\ref{sec: holdout}, uses the same $(\tau_{\mathrm{bug}}, \tau_{\mathrm{noise}}, \tau_{\mathrm{veto}}) = (0.35, 0.70, 0.20)$. Predictions in the intermediate range are flagged as \texttt{UNCERTAIN} for manual review. With the training data of Section~\ref{sec: training}, the model rarely uses the band: it abstains on no circuit of the primary suite and on one circuit of each held-out suite, in each case a bug whose divergence sits near the noise floor.

\subsection{Three-Class Classification with Uncertainty}
Rather than forcing binary predictions, the framework outputs three classes: \texttt{SOFTWARE\_BUG}, \texttt{HARDWARE\_NOISE}, and \texttt{UNCERTAIN}. The \texttt{UNCERTAIN} class represents samples where the classifier lacks sufficient confidence, following the selective classification framework of Geifman and El-Yaniv \cite{geifmanSelectiveClassificationDeep2017}.

\subsubsection{Decision Logic}
The classification proceeds in two stages. First, a physics-informed veto overrides the learned classifier when distributional divergence exceeds what hardware noise can plausibly produce. Second, the ANFIS probability output determines classification for remaining cases.

This three-zone classification reflects safety engineering principles articulated by Leveson \cite{levesonEngineeringSaferWorld2016}: in systems where incorrect decisions carry significant costs, explicitly modeling uncertainty is preferable to forcing binary choices. 

Given the ANFIS output $p=P(\mathrm{noise})\in[0,1]$ and the log-transformed Bhattacharyya distance $D_B^{\mathrm{log}}$:
\[
    \text{Classification} = \begin{cases} 
        \texttt{SOFTWARE\_BUG} & \text{if } D_B^{\mathrm{log}} > \tau_{\mathrm{veto}} \\
        \texttt{HARDWARE\_NOISE} & \text{else if } p > \tau_{\mathrm{noise}} \\
        \texttt{SOFTWARE\_BUG} & \text{else if } p < \tau_{\mathrm{bug}} \\
        \texttt{UNCERTAIN} & \text{otherwise}
    \end{cases}
\]

\begin{figure}
    \centering
    \includegraphics[width=0.5\linewidth]{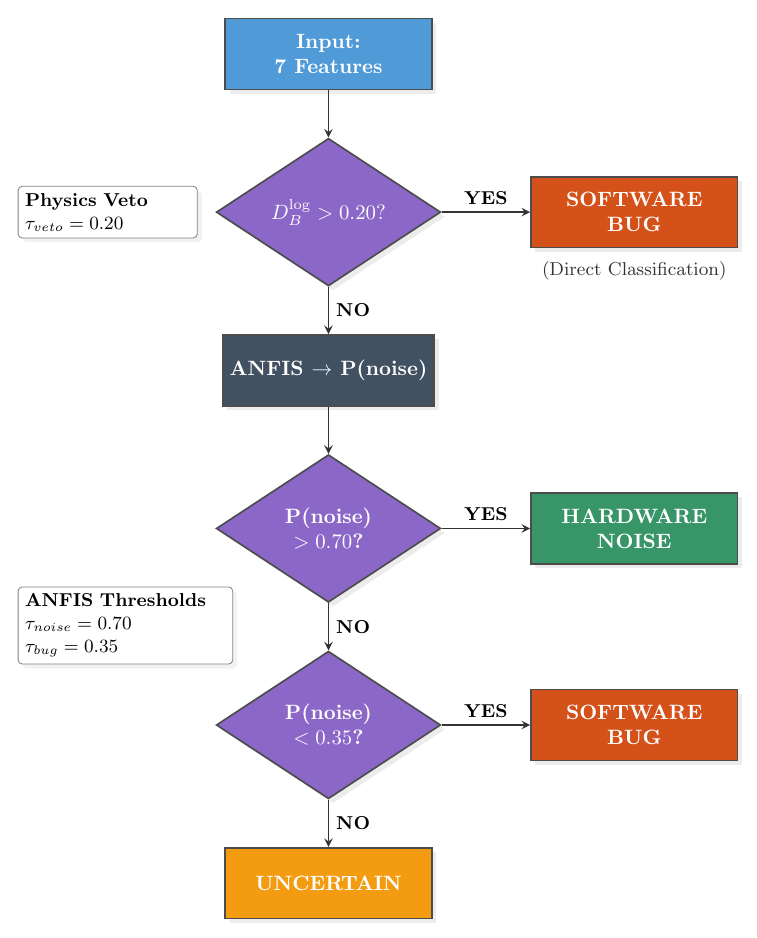}
    \caption{Classification Decision Logic: Veto check followed by ANFIS thresholds}
    \label{fig:flowchart}
\end{figure}

Classification thresholds $\tau_{\mathrm{bug}}=0.35$ and $\tau_{\mathrm{noise}}=0.70$ are fixed hyperparameters selected during development and not re-tuned for the model reported here (Section~\ref{sec: training}). The uncertainty zone spans 35\% of the probability space, requiring 65--70\% classifier confidence before committing to a definitive attribution. Figure \ref{fig:flowchart} illustrates the overall architecture.

\subsubsection{Physics-Informed Bhattacharyya Veto}\label{sec: veto}
Hardware noise on superconducting processors is well-modeled by completely
positive trace-preserving (CPTP) maps, including depolarizing channels,
amplitude damping, and dephasing. Such maps are contractive: the data
processing inequality gives
$D(\Lambda\rho_{\mathrm{bug}}, \Lambda\rho_{\mathrm{correct}}) \leq
D(\rho_{\mathrm{bug}}, \rho_{\mathrm{correct}})$ for any monotone distance $D$,
so the separation between a buggy and a correct execution observed on the same
noisy device can never exceed their noiseless separation. Contraction
constrains the veto in one direction only. It says nothing about the quantity
the veto thresholds, which compares a \emph{noisy} measured distribution with a
\emph{noiseless} ideal reference; and it cuts against completeness rather than
for it: a channel that drives the output toward the ideal distribution---depolarizing
noise on a circuit whose ideal output is near-uniform, for instance---reduces the
measured divergence, so a sufficiently noisy bug can fall below any threshold.
The veto is therefore a sufficient condition for a fault, not a necessary one,
and its false-positive rate must be established empirically.

We establish it as follows. Across the 61 correctly-executed circuits of the
\texttt{ibm\_fez} suite (ISA depth to 76) the maximum observed
$D_B^{\mathrm{log}}$ is $0.042$; across the 34 correctly-executed circuits of
the two \texttt{ibm\_kingston} held-out suites (ISA depth to 101) it is $0.057$,
excluding a single circuit at $0.246$ recorded while the backend reported a
\texttt{maintenance} status (the same circuit measured $0.052$ on the following
day under normal operation). Taking $f_0$ as the largest $D_B^{\mathrm{log}}$ among correct circuits of ISA depth $\le 22$, and $b$ as the upper-envelope slope of $f_0 + b\,\ln\!\left((1+d)/23\right)$ over the deeper circuits, gives $f_0 = 0.055$ and $b = 0.002$ across the 95 correct circuits, with the maintenance-status circuit the only point excluded: the floor is flat with depth over the range studied, and the eight correct circuits at ISA depth 60--101 span $0.005$--$0.057$, within the spread of the shallow ones. The Bhattacharyya distance separates these regimes
because of its geometric-mean form. The coefficient
\[
    BC = \sum_x \sqrt{p(x) \cdot q(x)}
\]
loses mass \emph{linearly} in the probability moved to states where the
ideal distribution assigns zero, since those terms vanish, and only at
\emph{second order} in probability redistributed among states the ideal
already occupies (Section~\ref{sec: resolution}). Hardware noise does both,
but leaks only a small fraction of mass off the ideal support---a few percent
on these devices---and that leak is what the measured floor quantifies. A bug
that relocates a large fraction of the mass to states outside the support
therefore separates from the floor by a factor of $3.5$ at the veto threshold
and by more than an order of magnitude for typical gate substitutions.

The constraint is implemented as a hard rule that precedes the ANFIS:
\[
    \text{IF } D_B^{\mathrm{log}} > \tau_{\mathrm{veto}} \text{ THEN classify as } \texttt{SOFTWARE\_BUG}
\]
with $\tau_{\mathrm{veto}} = 0.20$, corresponding to a Bhattacharyya
coefficient of approximately $0.80$ and sitting $3.5\times$ above the largest value observed on a correct circuit, once that outlier is excluded. We keep a single fixed threshold rather than a per-circuit adaptive bound because the measured floor does not require one; a depth-aware
form $\tau(d) = \max\!\left\{\tau_{\mathrm{veto}},\ 3.6\,\big(f_0 + b\max\{0, \ln((1+d)/23)\}\big)\right\}$ is provided in the replication package and changes no decision reported here.

Across every evaluation in this paper, the veto fired on 27 of 44 bugs on \texttt{ibm\_fez}, 11 of 13 on the development suite, and 10 of 13 on the test suite. It also fired on one correct circuit, the maintenance-status circuit above. With the training data of Section~\ref{sec: training}, the veto is largely redundant with the learned model: the ANFIS alone assigns $P(\mathrm{noise}) < \tau_{\mathrm{bug}}$ to 47 of those 48 bugs. The remaining one, the Grover overrotation fault, receives $P(\mathrm{noise}) = 0.44$, so without the veto it would have been flagged for review rather than missed. Section~\ref{sec: rule_analysis} shows that the model has learned the veto's antecedents as an explicit rule. We retain the veto as a
safety interlock whose behavior is verified rather than as a source of
accuracy: it is the component that would catch a future distribution shift
the classifier did not.

\subsubsection{Interpretability and Rule Extraction}
A key advantage of ANFIS over black-box neural networks is the ability to extract human-readable fuzzy rules. The trained model can be interpreted as a set of IF-THEN rules mapping feature combinations to classification tendencies. For example:

\begin{description}
    \item[Rule:] IF Entropy is \texttt{LOW} AND Bhattacharyya is \texttt{HIGH} $\to$ \texttt{BUG}
    \item[Interpretation:] The output distribution is pure (low entropy) but far from the target (high Bhattacharyya distance)---the physical signature of a unitary rotation error that coherently rotates the state to an incorrect region of Hilbert space.
\end{description}

This transparency supports quantum debugging workflows: developers receiving a \texttt{SOFTWARE\_BUG} attribution can inspect the firing rules to understand why the circuit was flagged, guiding debugging efforts toward the most likely error categories. Section~\ref{sec: rule_analysis} presents the five most influential rules extracted from our trained model along with their physical interpretations.

\section{EXPERIMENTAL SETUP}
The framework is evaluated on two IBM Heron r2 processors: a 105-circuit primary suite on \texttt{ibm\_fez}, on which the pipeline was developed, and two 30-circuit held-out suites on \texttt{ibm\_kingston}, one used once as a development check under each training distribution and one reserved as a test, scored once with every parameter frozen. This section details the hardware platforms, the circuit suites, the training-data generator, the evaluation protocol, and the training-pipeline faults uncovered during development.

\subsection{Hardware Platform}
All experiments were conducted on IBM's \texttt{ibm\_fez} backend, a 156-qubit Heron r2 processor accessed through IBM Quantum cloud services. The Heron architecture represents IBM's latest generation of superconducting quantum processors, featuring tunable couplers that significantly reduce crosstalk compared to the previous Eagle generation \cite{IBMQuantumDelivers, abughanemIBMQuantumComputers2025}.

At the time of the experiments (calibration snapshot of 13 December 2025), the
processor exhibited the following characteristics. We report medians across
the 156 qubits rather than means, because one qubit was non-operational and
reported a sentinel error value of $1.0$, which inflates the mean
single-qubit gate error by a factor of approximately 25.

\begin{itemize}
    \item $T_1$ relaxation time (median): $145.2 \, \mu\text{s}$
    \item $T_2$ dephasing time (median): $102.7 \, \mu\text{s}$
    \item Single-qubit gate error (SX, median): $2.7 \times 10^{-4}$
    \item Two-qubit gate error (CZ, median): $2.7 \times 10^{-3}$
    \item Qubit count: 156 (heavy-hexagonal topology)
\end{itemize}
These metrics represent typical operational conditions for current NISQ
devices. All circuits were executed with 4,096 measurement shots, giving a
standard error on an individual outcome probability of
$\sigma_p = \sqrt{p(1-p)/N} \leq 0.78\%$. Circuits were transpiled using
Qiskit's optimization level 3, which applies aggressive gate cancellation,
routing, and decomposition into the native gate set (CZ, ID, RZ, SX, X).
Circuit structural features---two-qubit gate density and depth---are
computed on the post-transpilation (ISA) circuit, so that they reflect the
operations actually executed on hardware rather than the abstract
pre-transpilation form. For the \texttt{ibm\_fez} suite, these two features
were recomputed by re-transpiling the logical circuits against \texttt{ibm\_fez} at optimization level~3 with a fixed seed, because the December execution's transpiled circuits were no longer retrievable; the five distribution features are taken directly from the December measurement counts. Re-transpilation may place circuits on different physical qubits than the December run, but depth and two-qubit
gate counts are stable under that choice.

\paragraph{Second device.} The held-out suites (Section~\ref{sec: holdout})
were executed on \texttt{ibm\_kingston}, a second 156-qubit Heron r2
processor, on 3--4 September 2026 with 4,096 shots and the same transpilation
settings. Its calibration at execution (medians across 156 qubits) was
$T_1 = 204.2\,\mu$s, $T_2 = 127.7\,\mu$s, and SX error $2.4 \times 10^{-4}$. The first held-out job was submitted while the backend reported a
\texttt{maintenance} status; the second ran under normal operation.

\subsection{Validation Circuit Suite}
A comprehensive validation suite of 105 circuits spanning 17 algorithm families was constructed. The suite was designed to cover diverse quantum computational patterns, qubit counts (2--5 qubits), and bug types. Table \ref{tab:circuit_families} outlines the 17 distinct algorithm families, specifying the circuit count and expected output characteristics for each.
\begin{table}[h]
  \caption{Validation Circuit Families}
  \label{tab:circuit_families}
  \begin{tabular}{lcl}
    \toprule
    Algorithm Family & Circuits & Expected Output Pattern \\
    \midrule
    Bell States & 14 & Two states with 50\% each \\
    GHZ States & 11 & Two states ($|0\rangle^{\otimes n}$ and $|1\rangle^{\otimes n}$) \\
    W States & 4 & $n$ states with Hamming weight 1 \\
    Cluster/Graph States & 4 & Uniform over computational basis \\
    QFT & 9 & Uniform for basis inputs; peaked for periodic \\
    Grover Search & 6 & Highly peaked at marked state(s) \\
    Deutsch-Jozsa & 4 & Deterministic $|0\rangle^{\otimes n}$ \\
    Bernstein-Vazirani & 4 & Deterministic secret string \\
    VQE Ansätze & 8 & Parameter-dependent superposition \\
    QAOA & 5 & Concentrated on low-energy states \\
    Teleportation (Dynamic) & 5 & State-dependent with feed-forward \\
    Superdense Coding & 6 & Deterministic classical message \\
    Phase Estimation & 5 & Peaked at eigenvalue encoding \\
    SWAP Test & 4 & State overlap estimation \\
    Clifford Circuits & 3 & Stabilizer state outputs \\
    Error Correction & 5 & Syndrome-dependent (Repetition codes) \\
    Hardware-Efficient Ansätze & 8 & Parameter-dependent distribution \\
    \bottomrule
  \end{tabular}
\end{table}

\subsubsection{Bug Injection}\label{sec: bug_injection}
For each circuit family, bugs were systematically injected following the taxonomy of Huang and Martonosi \cite{huangStatisticalAssertionsValidating2019a}:
\begin{itemize}
    \item \textbf{Missing Gate:} Removing a required gate (e.g., CNOT from Bell state preparation)
    \item \textbf{Wrong Gate:} Substituting an incorrect gate (e.g., CZ instead of CNOT)
    \item \textbf{Wrong Angle:} Incorrect rotation parameter (e.g., $\pi/4$ instead of $\pi/2$)
    \item \textbf{Wrong Target:} Gate applied to incorrect qubit(s)
    \item \textbf{Extra Gate:} Spurious gate addition (e.g., extra Z gate flipping phase)
\end{itemize}
The final validation suite comprises 61 correct circuits and 44 buggy circuits.

\subsection{Training Data Generation}\label{sec: training}
Training data was generated using Qiskit Aer's noise simulation. A noise model was constructed directly from \texttt{ibm\_fez} calibration data using \texttt{NoiseModel.from\_backend()}, capturing gate-dependent error rates, readout asymmetries, and qubit-specific $T_1/T_2$ times. To diversify the noise regime, 40\% of samples were instead drawn through synthetic depolarizing-plus-readout models at $1\times$, $1.5\times$, and $2\times$ the backend's median gate and readout error rates.

Correct circuits were drawn from 25 template families: 15 shallow families (Bell, GHZ, W, QFT, Grover, Deutsch--Jozsa, Bernstein--Vazirani, VQE and QAOA ansätze, and others), and ten deeper or deterministic-output families---W states to six qubits, GHZ states with alternative topologies, random Clifford circuits, Trotterized Ising evolution, inverse-QFT decoding, $n$-bit Bernstein--Vazirani, Toffoli full adders, three-qubit Grover with one or two marked states, hardware-efficient ansätze with two to four layers, and Simon's algorithm. Sampling is balanced between the two groups. Circuits span 2--6 qubits; after transpilation to the \texttt{ibm\_fez} target at optimization level~3, ISA depths have median 13, 95th percentile 96, and maximum 139, matching the range of the hardware suites.

Buggy circuits are produced by mutation-style injection (missing gate, wrong gate, wrong parameter, wrong target qubit) into a freshly drawn correct circuit, and are simulated \emph{under the same noise-source distribution as correct circuits}, so that the classifier learns to attribute a deviation rather than to detect the presence of noise. Every feature is computed against the intended circuit's ideal distribution and on the transpiled circuit, exactly as at inference. An injection whose ideal computational-basis distribution lies within $D_B^{\mathrm{log}} < 0.01$ of the intended circuit's is rejected: such a fault cannot be attributed from computational-basis counts by any method (Section~\ref{sec: error_analysis}), and labelling it a bug would only teach the model that noise-like outputs are bugs. Of 1,655 injections attempted, 655 (40\%) were rejected on this criterion---a rate worth noting for anyone using mutation testing to generate quantum bug corpora. The final set contains 1,983 samples (986 bug, 997 noise); a further 399-sample synthetic validation set drawn with a different seed and backend noise only is used for early stopping.

The ANFIS architecture employs 16 fuzzy rules, selected in preliminary experiments showing diminishing returns beyond this value. Training hyperparameters: learning rate 0.005, up to 500 epochs with early stopping.

Because the noise model is queried from the backend at generation time, the synthetic training distribution depends on the device calibration state on that date; a fixed random seed does not by itself make the pipeline reproducible. We therefore archive the generated feature arrays alongside the generation script in the replication package.

\subsection{Evaluation Metrics}
Classifier performance was evaluated using metrics appropriate for three-class classification with rejection \cite{geifmanSelectiveClassificationDeep2017}:

\textbf{Effective Accuracy} counts both correct predictions and uncertain predictions as ``safe'':
\[
    \mathrm{Acc}_{\mathrm{eff}} = \frac{\mathrm{Correct} + \mathrm{Uncertain}}{\mathrm{Total}}
\]
This metric reflects that flagging a circuit for manual review is preferable to making an incorrect attribution. Effective accuracy must be read jointly with the abstention rate: a classifier that abstains on every input attains $100\%$ effective and $0\%$ strict accuracy. We therefore report coverage and selective accuracy alongside it throughout; for the model reported here the three figures coincide on the primary suite because it abstains on no circuit.

\textbf{Strict Accuracy} counts only definitive correct predictions:
\[
    \mathrm{Acc}_{\mathrm{strict}} = \frac{\mathrm{Correct}}{\mathrm{Total}}
\]

\textbf{Uncertainty Rate} measures the proportion of circuits flagged for manual review:
\[
    U = \frac{\mathrm{Uncertain}}{\mathrm{Total}}
\]
Lower uncertainty is desirable, but not at the cost of increased errors. 

We also report 95\% Wilson confidence intervals \cite{wilsonProbableInferenceLaw1927} for all proportions, accounting for the finite validation set size.

\subsection{Held-Out Evaluation Protocol}\label{sec: protocol}
The \texttt{ibm\_fez} suite served, during development, as the set on which classification thresholds were selected, the veto floor characterized, and rules ranked; a result on it alone cannot separate generalization from tuning. We therefore constructed a 30-circuit held-out suite of 17 correct and 13 buggy circuits (approximately the 61:44 ratio of the primary suite) from families and instances absent from the primary suite: Simon's algorithm, five-qubit W states, half adders, an eight-outcome random Clifford circuit on five qubits, Trotterized Ising evolution, two-marked-state Grover search, four-bit Bernstein--Vazirani and Deutsch--Jozsa, a six-qubit GHZ state, inverse-QFT decoding, a five-qubit hardware-efficient ansatz, and Bell states measured in the X basis. Bugs were injected by hand following the taxonomy of Section~\ref{sec: bug_injection} and include three deliberately hard cases whose ideal divergence lies below the veto (seed-0 values: a W-state rotation error at $D_B^{\mathrm{log}} = 0.05$, a Trotter angle error at $0.09$, and an inverse-QFT phase error at $0.17$). The suite is also deeper than the primary one: ISA depths of the correct circuits have median 27 and maximum 101 in seed~0, and 100 in seed~2, against 15 and 76 for \texttt{ibm\_fez}.

Two instances of the suite were executed on \texttt{ibm\_kingston}. The first (seed~0) was run against a model trained on the shallow pipeline described in Section~\ref{sec: corrections}, exposed the failure mode analyzed there, and was re-scored once under the training distribution of Section~\ref{sec: training}; it is reported as a \emph{development} result. The second (seed~2) draws fresh instances of every family---different secret strings, marked states, rotation targets, Clifford seeds, and register widths---and was executed and scored exactly once, after every parameter of the pipeline had been frozen. It is the \emph{held-out test} result, and no number in this paper was changed after it was seen.

\subsection{Requirements on the Training Distribution}\label{sec: corrections}

Three properties proved necessary for a synthetic training set to transfer effectively to hardware. An earlier build of the generator, hereafter the shallow pipeline, satisfied none of them. First, structural features must be computed on the transpiled circuit: if taken from the logical circuit, a three-qubit Grover built from Toffoli gates reports a two-qubit gate density of zero, and hardware depths reaching 100 fall outside the training range, where the Gaussian memberships of every rule vanish. Second, every sample must carry device noise, or the absence of noise becomes a bug signature. Third,
injected bugs whose ideal output is indistinguishable from the correct circuit must be filtered---40\% of mutation-style injections in our generator---or the model is trained to attribute them. Evaluated on the seed-0 suite, the shallow pipeline reaches 70.0\% strict accuracy against 93.3\% for the pipeline of Section~\ref{sec: training} (Table~\ref{tab:holdout}). All five of its errors are correct circuits attributed to bugs---Simon's algorithm, two inverse-QFT circuits, a two-marked Grover search, and the maintenance-status Clifford circuit---and three of its four abstentions are correct circuits as well. Neither failure mode appears in the results reported here.

\section{RESULTS AND ANALYSIS}\label{sec: results}

This section presents the results on the \texttt{ibm\_fez} primary suite, analyzes the learned rule base, examines the residual errors, and then reports the held-out evaluation on \texttt{ibm\_kingston}.

\subsection{Overall Performance}

\begin{figure}[h]
    \centering
    \includegraphics[width=0.85\linewidth]{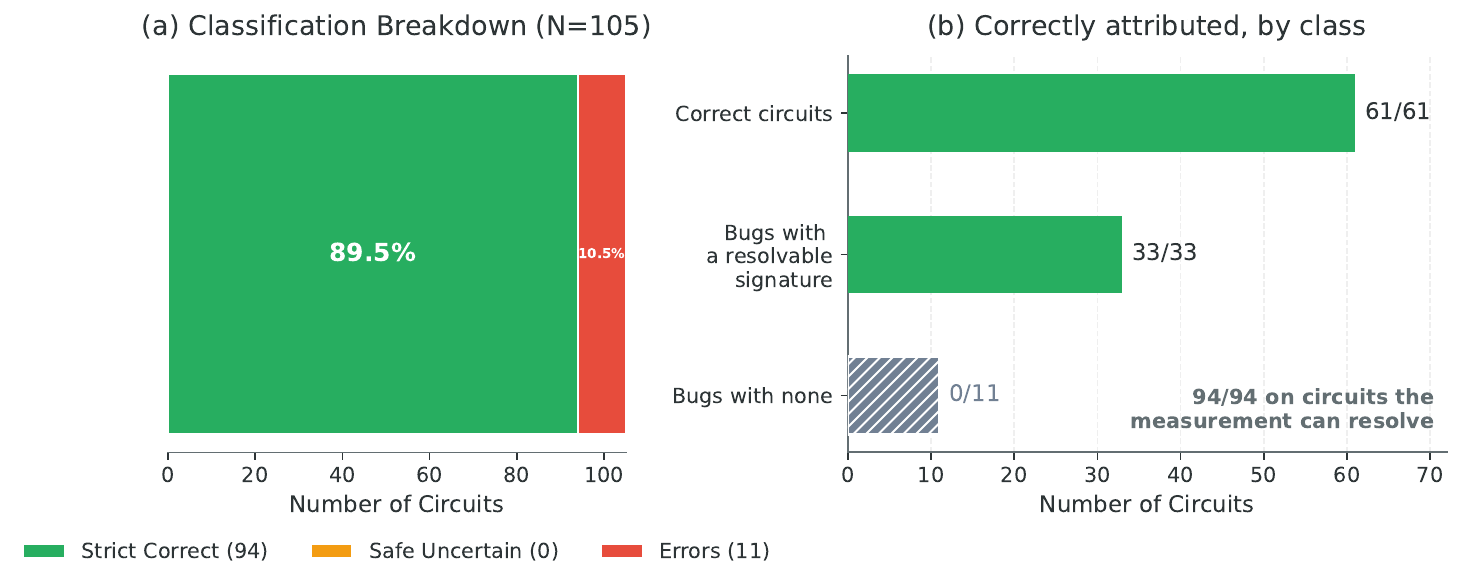}
    \caption{Classification outcomes on the 105-circuit primary suite. (a) Overall: 94 correct, no abstentions, 11 errors. (b) By class: every correct circuit is attributed correctly. Nine of the 11 errors are bugs with no computational-basis signature; the other two shift the output by less than the noise floor of the divergence feature (Section~\ref{sec: error_analysis}).}
    \label{fig:classification_results}
\end{figure}

Table~\ref{tab:results} summarizes the classification results on the 105-circuit primary suite.
\begin{table}[h]
\centering
\caption{Classification Results on IBM Heron r2 (\texttt{ibm\_fez}), 105-circuit primary suite}
\label{tab:results}
\begin{tabular}{lll}
    \toprule
    \textbf{Metric} & \textbf{Value} & \textbf{Context} \\
    \midrule
    Total Circuits & 105 & 61 correct, 44 buggy \\
    Strict Correct & 94 (89.5\%) & Exact match with ground truth \\
    Safe Uncertain & 0 (0.0\%) & Flagged for human review \\
    Errors & 11 (10.5\%) & All bugs attributed to noise; see Table~\ref{tab:error_categories} \\
    Strict Accuracy & 89.5\% & Wilson 95\% CI $[82.2\%,\, 94.0\%]$ \\
    Coverage & 100\% & Effective and selective accuracy coincide at 89.5\% \\
    Correct circuits & 61/61 & No correct circuit attributed to a bug \\
    Detectable bugs & 33/35 & Nine of the 44 are Z-basis invisible (Section~\ref{sec: error_analysis}) \\
    Shots per Circuit & 4,096 & Shot noise $\sigma_p \leq 0.78\%$ \\
    \bottomrule
\end{tabular}
\end{table}
The model commits on every circuit and is correct on 94 of them. All 11 errors are buggy circuits attributed to hardware noise, and Section~\ref{sec: error_analysis} shows that nine of these have ideal computational-basis distributions identical to their intended circuits---no method operating on Z-basis counts could separate them---while the remaining two lie below the device's divergence floor. Relative to those limits, the model resolves 94 of the 96 circuits that are resolvable. Figure~\ref{fig:classification_results} shows the breakdown.

\begin{figure}
    \centering
    \includegraphics[width=0.8\linewidth]{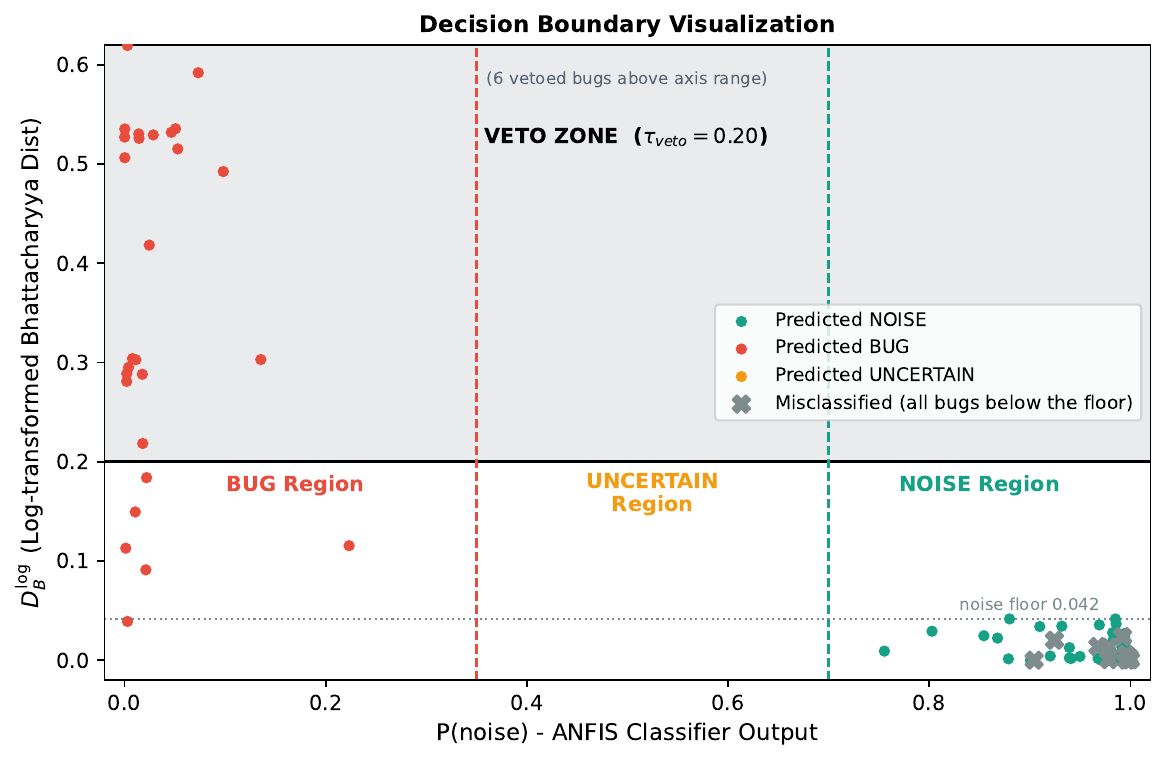}
    \caption{Decision logic on the 105-circuit primary suite. The shaded band
    above $D_B^{\mathrm{log}} = 0.20$ is the veto region; below it, the boundaries
    at $P(\mathrm{noise}) = 0.35$ and $0.70$ separate bug, uncertain, and noise
    regions. Dotted line: measured noise floor ($0.042$). The axis is clipped at
    $0.62$; six vetoed bugs lie above it. Crosses mark misclassifications, all
    below the floor.}
    \label{fig:decision_boundary}
\end{figure}

Figure~\ref{fig:decision_boundary} maps the decision logic onto the feature space. The horizontal band above $D_B^{\mathrm{log}} = 0.20$ is the veto region; 27 circuits fall in it, all of them buggy. Below the veto threshold, the vertical boundaries at $P(\mathrm{noise}) = 0.35$ and $0.70$ partition the space into bug, uncertain, and noise regions. No circuit of the primary suite is abstained: correct circuits cluster at $P(\mathrm{noise})$ near 1 and detectable bugs near 0, with the 11 missed bugs sitting among the correct circuits at $D_B^{\mathrm{log}} \le 0.024$.

\subsection{The Grover Boundary Problem: A Success Case}\label{sec:grover}
A key validation of our approach is the correct handling of Grover's algorithm circuits. The framework classifies \texttt{grover\_2q\_correct} as \texttt{HARDWARE\_NOISE} despite its highly peaked output distribution ($\text{bias} = 0.821$), and also correctly resolves \texttt{grover\_2q\_2iter\_correct} and \texttt{grover\_3q\_correct}, whose two-iteration and three-qubit structure produce deeper transpiled circuits.

The entropy deviation feature ($\Delta H = 0.297$ for \texttt{grover\_2q\_correct}) is key. The expected entropy for a correctly-implemented Grover search is low---the algorithm should produce a peaked distribution---so the small measured-to-expected entropy difference indicates correct operation despite noise. A buggy circuit producing an equally peaked distribution would show high entropy deviation because its expected entropy would differ. Section~\ref{sec: rule_analysis} shows that the trained model encodes this resolution as an explicit rule: low entropy and high bias with low entropy deviation conclude \texttt{NOISE}.

\begin{figure}
    \centering
    \includegraphics[width=0.9\linewidth]{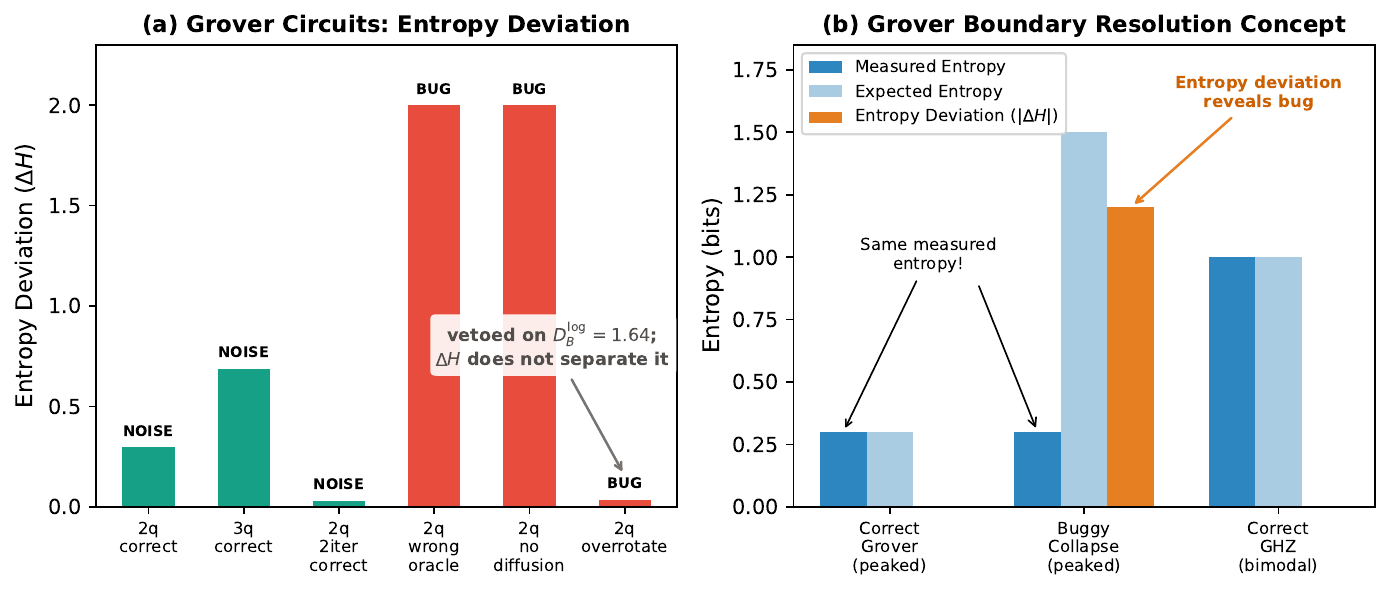}
    \caption{Resolution of the Grover boundary problem. (a) Grover circuits from
the primary suite, colored by attribution; all six are correctly classified.
(b) Conceptual illustration with representative values, not measurements.}
    \label{fig:grover_boundary}
\end{figure}

Figure~\ref{fig:grover_boundary} illustrates the mechanism. The left panel shows the Grover circuits of the primary suite; the right panel the underlying concept: both a correctly-functioning Grover search and a buggy ``collapse'' circuit produce low measured entropy (peaked outputs), but only the buggy circuit shows high entropy deviation because its expected output should have been a superposition. This differential diagnostic is what allows the framework to distinguish ``peaked by design'' (Grover, Bernstein--Vazirani, Deutsch--Jozsa) from ``peaked by accident'' due to software faults. Lower entropy indicates
a more concentrated output, so the two peaked cases share a measured value and
differ only in what was expected. The overrotation fault is the exception the
left panel makes visible: its entropy deviation is smaller than that of two
correct circuits, and it is caught by the Bhattacharyya veto rather than by
$\Delta H$.

\subsection{Successful Detection Patterns}
The framework detects every bug whose ideal output differs from the intended one by more than the noise floor. Three patterns dominate:

\begin{description}
    \item[Grover oracle bugs.] The \texttt{grover\_2q\_wrong\_oracle} circuit
produced $\Delta H = 1.999$ and $D_B^{\mathrm{log}} = 0.535$, triggering
bug classification via both the entropy and veto pathways.
    \item[Missing entanglement bugs.] Circuits with missing CNOT gates in GHZ preparation showed entropy deviation between $0.18$ and $0.41$, well above that of the correct GHZ circuits, because the expected GHZ state $(|000\dots\rangle + |111\dots\rangle)/\sqrt{2}$ collapsed to a product state with different entropy characteristics.
    \item[Wrong gate substitutions.] Substituting X for H consistently produced Bhattacharyya distances exceeding the veto threshold, as these substitutions rotate states to measurably different output distributions.
\end{description}

\begin{figure}
    \centering
    \includegraphics[width=0.6\linewidth]{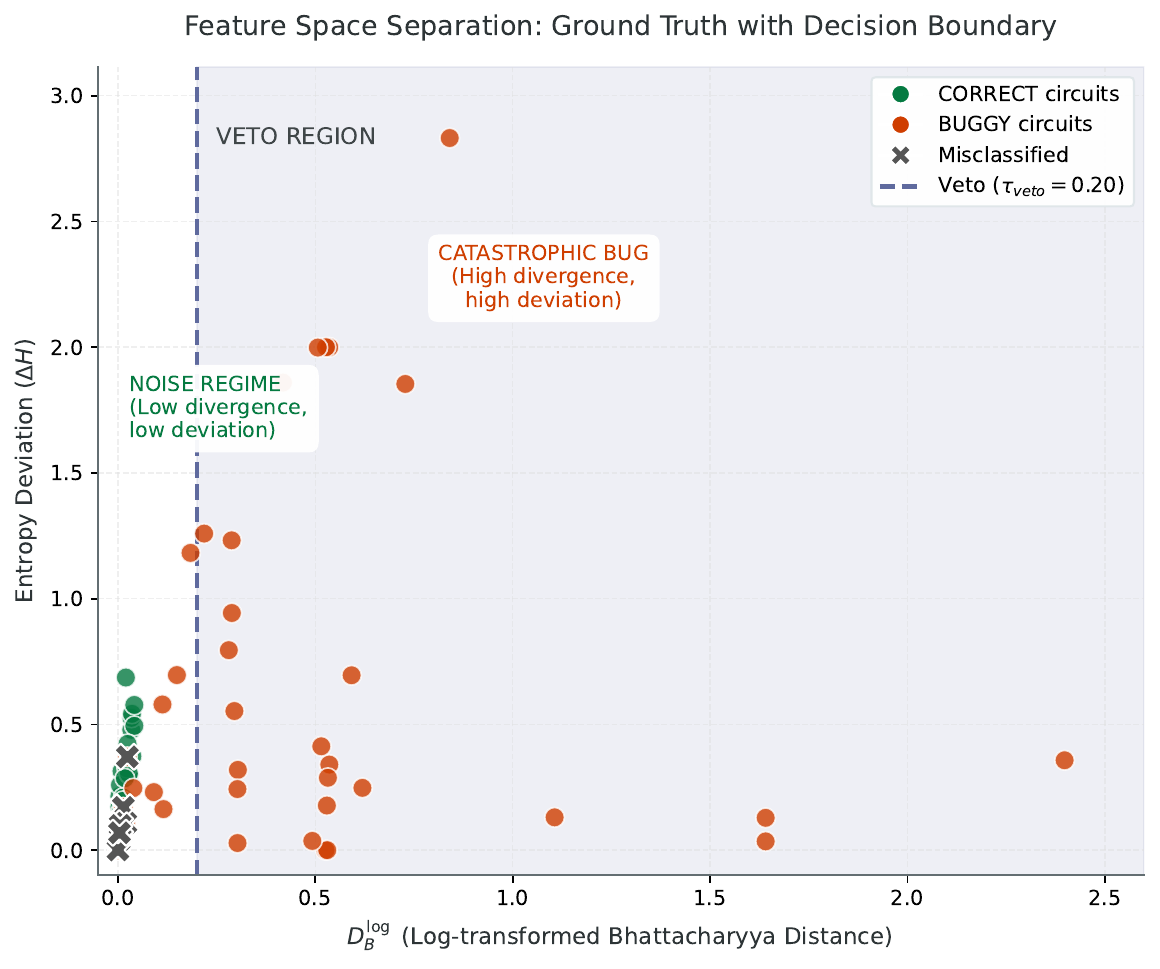}
    \caption{Feature space visualization of log-transformed Bhattacharyya distance ($D_B^{\mathrm{log}}$) versus entropy deviation ($\Delta H$) for all 105 primary-suite circuits. Green: correctly-executed circuits; red: buggy circuits; X: misclassified. Dashed line: veto threshold at $D_B^{\mathrm{log}} = 0.20$. The misclassified bugs lie within the correct-circuit cluster.}
    \label{fig:feature_space}
\end{figure}

Figure \ref{fig:feature_space} visualizes the two most discriminative features---log-transformed Bhattacharyya distance and entropy deviation---across all 105 circuits. The vertical dashed line at $D_B^{\mathrm{log}}=0.20$ marks the veto threshold. Correct circuits cluster in the lower-left; detectable bugs scatter toward higher values; the misclassified circuits, marked with X, sit inside the correct cluster: for nine of them the measured distribution is the correct distribution plus noise, and the other two differ from it by less than the noise floor.

\begin{figure}[h]
    \centering
    \includegraphics[width=0.9\linewidth]{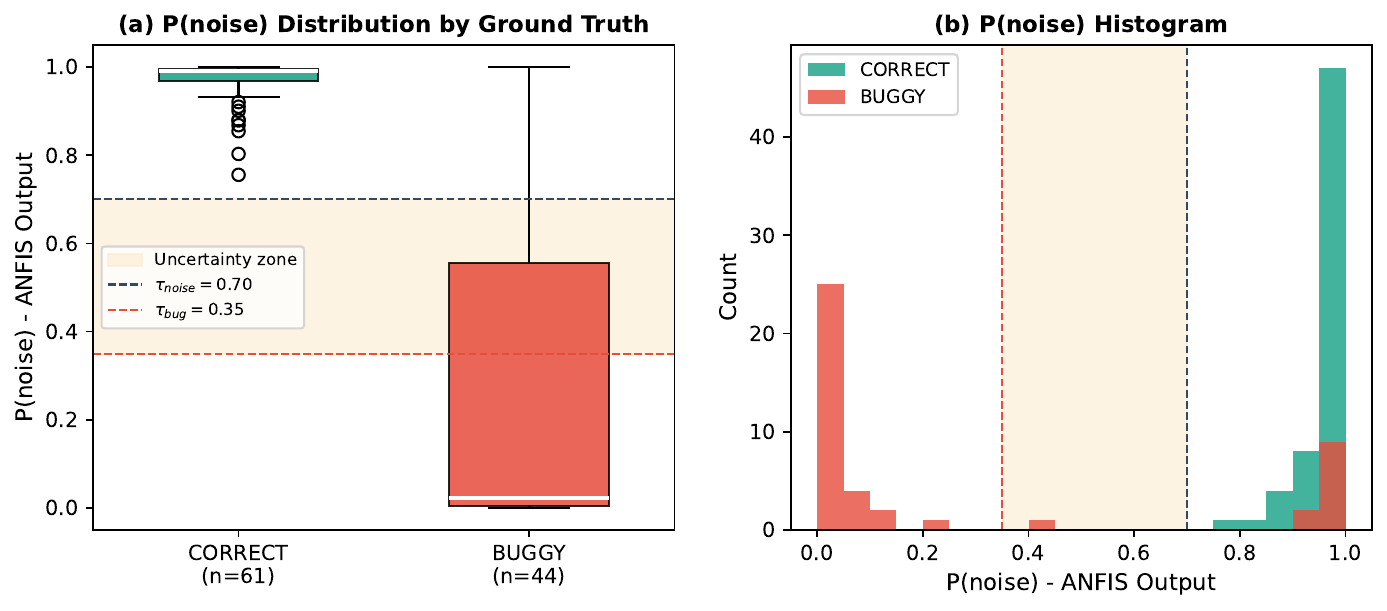}
    \caption{Distribution of ANFIS output $P(\mathrm{noise})$ by ground-truth class on the primary suite. Correct circuits cluster near 1 and buggy circuits near 0. One bug falls in the yellow-shaded uncertainty zone (0.35--0.70): the Grover overrotation fault, which the veto classifies first. The buggy circuits at high $P(\mathrm{noise})$ are the Z-basis-invisible and below-floor faults of Table~\ref{tab:error_categories}.}
    \label{fig:pnoise_distribution}
\end{figure}

Figure~\ref{fig:pnoise_distribution} shows how the ANFIS output $P(\mathrm{noise})$ separates the two ground-truth classes. Correct circuits cluster near $P(\mathrm{noise}) = 1$ and detectable bugs near $0$. One circuit falls inside the 0.35--0.70 band: the Grover overrotation fault, at $P(\mathrm{noise}) = 0.44$. The veto classifies it before the band is consulted ($D_B^{\mathrm{log}} = 1.64$), so no primary-suite circuit is abstained. The bimodal structure indicates that the model has learned a decisive representation. The band decides a case only on the held-out suites (Section~\ref{sec: holdout}), and there it falls on a bug whose divergence sits near the noise floor.

\subsection{Extracted Fuzzy Rules and Physical Interpretation}\label{sec: rule_analysis}

A central advantage of ANFIS over opaque classifiers is the ability to inspect the rules the model has learned. We converted the 16 trained fuzzy rules into human-readable IF-THEN statements: each rule's antecedents come from the centers of its Gaussian membership functions (LOW for normalized centers below $-0.5$, HIGH above $+0.5$, MEDIUM in between, and omitted when within that band), and its conclusion comes from the sigmoid of its consequent bias term, applying the same thresholds used at inference ($P(\mathrm{noise}) < 0.35 \to$ BUG, $> 0.70 \to$ NOISE, otherwise UNCERTAIN). Rules were ranked by mean normalized firing strength across the 105 primary-suite circuits.

Table~\ref{tab:extracted_rules} presents the five most influential rules, which together account for 68.9\% of the cumulative firing strength; the next four (R5, R11, R7, R14) bring the total to 94.5\%, and six rules fire on essentially no hardware circuit. Rules are indexed from 1 following Jang's convention \cite{jangANFISAdaptivenetworkbasedFuzzy1993}, matching the $w_i$ labels of Figure~\ref{Architecture}.

\begin{table*}[!t]
\centering
\caption{Top Five Extracted Fuzzy Rules, Ranked by Mean Firing Strength on the 105 Primary-Suite Circuits. Firing is each rule's share of total normalized firing strength across the 105 circuits; the conclusion's number is the sigmoid of the rule's consequent bias.}
\label{tab:extracted_rules}
\small
\renewcommand{\arraystretch}{1.3}
\begin{tabularx}{\textwidth}{@{}
    >{\centering\arraybackslash}p{0.04\textwidth}
    >{\centering\arraybackslash}p{0.07\textwidth}
    >{\raggedright\arraybackslash}p{0.27\textwidth}
    >{\centering\arraybackslash}p{0.13\textwidth}
    >{\raggedright\arraybackslash}X
@{}}
\toprule
\textbf{Rule} & \textbf{Firing} & \textbf{Antecedents (IF\ldots)} & \textbf{Conclusion} & \textbf{Physical Interpretation} \\
\midrule
R13 & 29.8\% & $p_{\max}$ LOW, depth LOW, $\Delta H$ LOW, $D_B^{\log}$ LOW & NOISE ($0.73$) & The base noise rule. A shallow circuit whose output is spread (no dominant outcome), matches its expected entropy, and stays close to the ideal distribution is what a correct circuit under weak decoherence looks like. \\
\addlinespace
R12 & 12.2\% & $H$ LOW, $\beta$ HIGH, $p_{\max}$ HIGH, depth LOW, $D_B^{\log}$ HIGH & BUG ($0.21$) & Sharply peaked output that is far from the target: probability has been concentrated on the wrong basis state. This is the ``low entropy, high distance'' signature of a unitary error that Section~\ref{sec: veto} describes---collapse to an incorrect deterministic outcome. \\
\addlinespace
R1  & 10.4\% & $p_{\max}$ LOW, depth HIGH, 2Q HIGH, $\Delta H$ LOW, $D_B^{\log}$ LOW & UNCERTAIN ($0.68$) & A deep, entanglement-heavy circuit with spread output and low divergence. The rule leans toward noise but its bias alone stops short of the threshold, deferring to its consequent slopes; together with R11 it covers the depth regime a shallow training distribution never reaches. \\
\addlinespace
R6  & 9.2\% & $H$ LOW, $\beta$ HIGH, depth LOW, $\Delta H$ LOW, $D_B^{\log}$ LOW & NOISE ($0.76$) & Peaked by design. Low entropy and high bias would suggest collapse, but the entropy deviation is small and the divergence low: the output is peaked because the algorithm intended it (Grover, Bernstein--Vazirani, Deutsch--Jozsa). This is the Grover-boundary resolution of Section~\ref{sec:grover}, learned as a rule. \\
\addlinespace
R9  & 7.3\% & $\beta$ LOW, depth LOW, 2Q LOW, $\Delta H$ HIGH, $D_B^{\log}$ HIGH & BUG ($0.24$) & Large entropy deviation and topological divergence on a shallow, lightly entangled circuit. The limited noise budget cannot account for the discrepancy---the canonical coherent gate-substitution fingerprint. This rule reproduces the Bhattacharyya-veto logic and retained nearly identical antecedents across retraining. \\
\bottomrule
\end{tabularx}
\end{table*}

Three observations emerge. First, the rule base contains explicit rules for the two regimes that a shallow training distribution leaves uncovered: R6 encodes the Grover-boundary resolution (peaked but expected), and R1 together with R11 (6.3\%; depth HIGH, 2Q HIGH, $D_B^{\log}$ LOW $\to$ NOISE at $0.84$) encode that deep, entangled circuits with low divergence are noisy rather than buggy. Neither rule existed in the rule base of the model trained on the shallow pipeline of Section~\ref{sec: corrections}. Second, R9 independently reproduces the physics of the Bhattacharyya veto---high divergence combined with a small noise budget---and did so in both models with nearly the same antecedents, which is the strongest evidence in this paper that the veto encodes something the data also contain. Third, the two highest-firing rules conclude in opposite directions (R13 toward NOISE, R12 toward BUG) and both sit outside the abstention band, so the model's confidence on the primary suite comes from the rules that fire most, not from a few sharply separated minor rules.

\FloatBarrier
\subsection{Error Analysis} \label{sec: error_analysis}
The residual errors reveal the constraints on what can be inferred from single-basis measurement statistics. In quantum detection theory, the minimum probability of error in distinguishing two quantum states is given by the Helstrom bound, which depends on the trace distance between states \cite{helstromQuantumDetectionEstimation1969}. When measurement is restricted to the computational (Z) basis, this bound applies to the classical distributions obtained by projection. When those projected distributions are identical, the Helstrom error probability is exactly $0.5$---random guessing---and no classifier operating on the counts can do better.

For each of the 11 misclassified circuits, we simulated the buggy and the intended circuit without noise and computed the divergence between their ideal computational-basis distributions. Table~\ref{tab:error_categories} gives the result: in nine cases that divergence is exactly zero, and in the remaining two it is below the device's measured noise floor.
\begin{table}[h]
\centering
\caption{Classification Errors by Category (primary suite). All 11 errors are buggy circuits attributed to noise; no correct circuit was attributed to a bug.}
\label{tab:error_categories}
\begin{tabular}{lcc>{\raggedright\arraybackslash}p{0.49\linewidth}}
    \toprule
    \textbf{Category} & \textbf{Count} & \textbf{Fraction} & \textbf{Root Cause} \\
    \midrule
    Z-basis invisible & 9 & 82\% & Ideal computational-basis distribution identical to the intended circuit's ($D_B = 0$ in noiseless simulation). No classifier can separate these from the counts. \\
    Divergence floor limit & 2 & 18\% & Shift present, but its Bhattacharyya distance is below the device's measured floor. One of the two is recovered by a decision-tree baseline that uses the same features (Section~\ref{sec: baselines}). \\
    \bottomrule
\end{tabular}
\end{table}

\subsubsection{Z-Basis Invisible Faults (9 errors, 82\%)}
\begin{description}
    \item[Circuits:] \texttt{bell\_extra\_z\_buggy}, \texttt{cluster\_star\_missing\_cz}, \texttt{qft\_2q\_missing\_cp}, \texttt{qft\_3q\_missing\_cp}, \texttt{qft\_4q\_missing\_cp}, \texttt{swap\_test\_missing\_h}, \texttt{bitflip\_incomplete}, \texttt{superdense\_11\_incomplete}, \texttt{error\_detect\_incomplete}
\end{description}
These circuits produce, in noiseless simulation, output distributions identical to their intended counterparts when measured in the computational basis, either because the injected fault changes only relative phases or because the removed gate acts trivially on the tested input state. For \texttt{bell\_extra\_z\_buggy}, the bug inserts an extra Z gate, converting $|\Phi^+\rangle = (|00\rangle + |11\rangle)/\sqrt{2}$ to $|\Phi^-\rangle = (|00\rangle - |11\rangle)/\sqrt{2}$; both yield $P(00) = P(11) = 0.5$. The measured $D_B^{\mathrm{log}}$ values of these nine circuits ($0.0001$ to $0.024$) are noise, and the classifier's confident \texttt{HARDWARE\_NOISE} attributions ($P(\mathrm{noise}) \ge 0.90$) are, given the counts, correct. That this category dominates the residual errors is a property of the bug corpus rather than of the model: mutation-style injection produces such faults at a high rate (40\% of injections in our generator; Section~\ref{sec: training}), and a corpus built for attribution studies should either filter them or report them separately, as we do. The held-out suite shows the remedy directly: the same extra-Z fault on a Bell pair measured in the X basis produced $D_B^{\mathrm{log}} = 1.136$ in the test suite and was vetoed (Section~\ref{sec: holdout}).

\subsubsection{Divergence Floor Limit (2 errors, 18\%)}
\begin{description}
    \item[Circuits:] \texttt{bell\_angle\_5deg\_buggy}, \texttt{bell\_angle\_15deg\_buggy}
\end{description}
These circuits are constructed as $R_Y(\pi/2 + \epsilon)$ followed by
CNOT, so the ideal output probability is $P(00) = (1 - \sin\epsilon)/2$ and
the injected fault shifts probability by $\delta = \sin(\epsilon)/2$: $4.36$
percentage points at $\epsilon = 5^\circ$ and $12.94$ points at $15^\circ$.
At $N = 4096$ shots these are $5.6\sigma$ and $16.6\sigma$ effects
respectively, so the faults are not obscured by sampling noise. They are
obscured by the feature.

For a redistribution of probability between two equiprobable outcomes, the
Bhattacharyya coefficient expands as $BC \approx 1 - \delta^2/2$ (the general
form is given in Section~\ref{sec: resolution}), so $D_B$ grows
\emph{quadratically} in the probability shift rather than in proportion to it.
A shift of $4.36$ points therefore produces $D_B^{\mathrm{log}} \approx 0.001$.
Adding the measured noise baseline of the correct Bell circuit
($D_B^{\mathrm{log}} = 0.0114$) predicts approximately $0.0124$ and $0.0200$
for the two faults, against measured values of $0.0139$ and $0.0202$---agreement
to within the sampling spread, treating the two contributions as additive
and neglecting their cross term. Both fall far below the veto threshold and
present to the ANFIS as noise-level divergence ($P(\mathrm{noise}) = 0.97$ and
$0.92$).

The limitation is therefore a property of how the divergence scores the shift
rather than of single-basis measurement or of the shot budget. Solving the
exact Bhattacharyya expression (not its quadratic expansion) for the bug
contribution alone against the empirical floor of $0.042$ gives
$\delta > 0.275$, i.e.\ $\epsilon > 33^\circ$ for this circuit family;
including the Bell circuit's own baseline places the crossing near
$28^\circ$. Section~\ref{sec: resolution} develops this bound.

This limit belongs to the Bhattacharyya feature, not to the full feature set. The decision-tree baseline of Section~\ref{sec: baselines}, trained on the same data and features, attributes the $15^\circ$ fault correctly, and it does so through entropy deviation. When hardware noise raises a circuit's divergence, it usually also raises its entropy, because it spreads probability onto outcomes the ideal output does not use. The $15^\circ$ fault raises the divergence but lowers the entropy, because it moves probability toward one of the two intended outcomes. Its entropy deviation ($0.107$) is therefore smaller than that of the correct Bell circuit ($0.154$). ANFIS does not weight this pattern strongly enough to change its decision and assigns the fault $P(\mathrm{noise}) = 0.92$. Section~\ref{sec: resolution} shows that the pattern holds on both devices.

\subsection{Held-Out Evaluation on a Second Device}\label{sec: holdout}
Table~\ref{tab:holdout} reports the two 30-circuit suites of Section~\ref{sec: protocol}, executed on \texttt{ibm\_kingston} and scored with every threshold frozen at the values of Section~\ref{sec: training}.

\begin{table}[h]
\centering
\caption{Held-out evaluation on \texttt{ibm\_kingston} (30 circuits each: 17
correct, 13 buggy). Seed~0 was scored once under each training distribution and is a
development result; seed~2 was executed and scored once with all parameters
frozen and is the test result.}
\label{tab:holdout}
\small
\begin{tabular}{lcccccc}
\toprule
\textbf{Suite} & \textbf{Model} & \textbf{Strict} & \textbf{Abstain} & \textbf{Errors} & \textbf{Effective} & \textbf{Strict (Wilson 95\%)} \\
\midrule
Seed 0 (dev.) & shallow pipeline & 21 & 4 & 5 & 83.3\% & 70.0\% $[52.1, 83.3]$ \\
Seed 0 (dev.) & this work & 28 & 1 & 1 & 96.7\% & 93.3\% $[78.7, 98.2]$ \\
Seed 2 (test) & this work & 28 & 1 & 1 & 96.7\% & 93.3\% $[78.7, 98.2]$ \\
\bottomrule
\end{tabular}
\end{table}

\begin{figure}[h]
    \centering
    \includegraphics[width=0.8\linewidth]{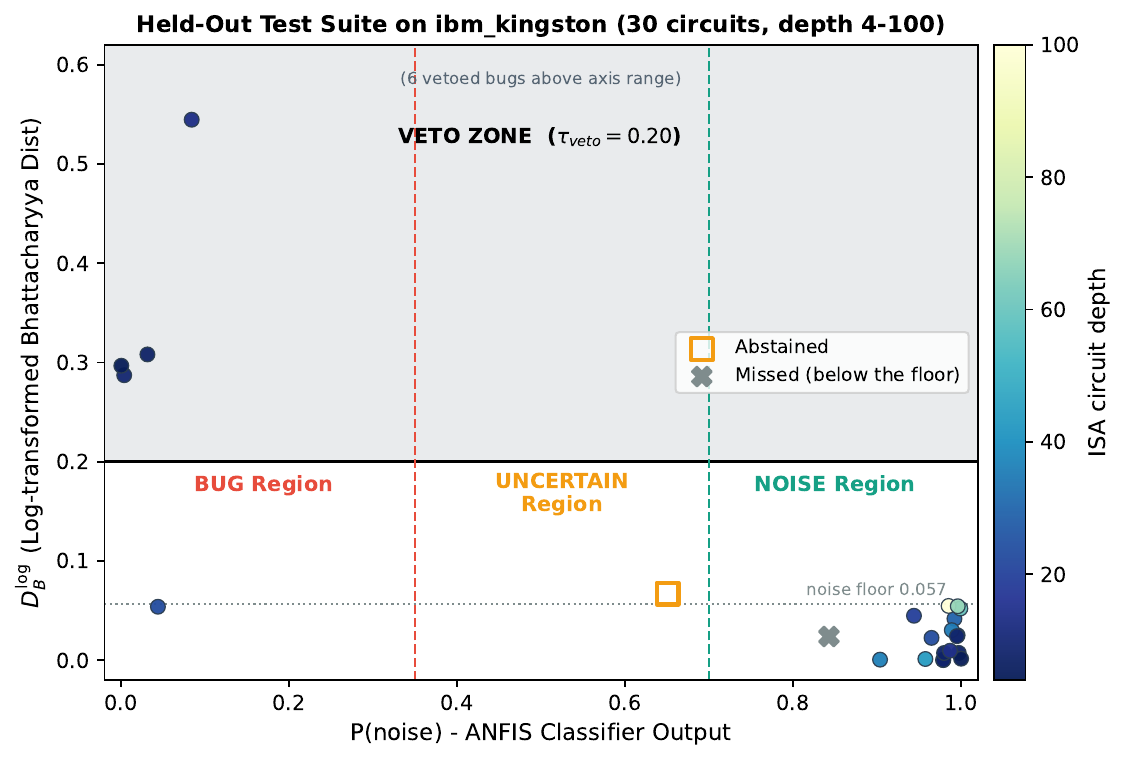}
    \caption{Held-out test suite on \texttt{ibm\_kingston}, scored with all
thresholds frozen. Color indicates ISA depth; the dotted line is this device's
noise floor ($0.057$). The axis is clipped at $0.62$, above which six vetoed
bugs lie. The square marks the abstention, the cross the miss.}
    \label{fig:holdout_boundary}
\end{figure}

Figure~\ref{fig:holdout_boundary} maps the held-out circuits onto the same
decision space as Figure~\ref{fig:decision_boundary}. Depth, which reaches 100
against the primary suite's maximum of 76, is not associated with degraded
confidence: the three deepest correct circuits are attributed to noise with $P(\mathrm{noise}) \ge 0.985$. 

On the held-out test suite all 17 correct circuits are attributed to noise with $P(\mathrm{noise}) \ge 0.90$, including the depth-100 two-marked Grover circuit, the six-qubit GHZ state, and the deterministic-output adders, Bernstein--Vazirani, Deutsch--Jozsa, and inverse-QFT circuits---the class of circuit on which a model trained on the shallow pipeline of Section~\ref{sec: corrections} failed. Of the 13 bugs, 11 are attributed to bugs (10 by the veto, one by the ANFIS below it), one is abstained, and one is missed. The abstention is the Trotter angle error, whose measured divergence ($0.067$) sits just above the noise floor of $0.057$ and far below the veto; the miss is a rotation error in the hardware-efficient ansatz whose \emph{ideal} divergence from the intended circuit is $0.019$, below the floor of $0.057$. It is a divergence-floor case of the kind analyzed in Section~\ref{sec: error_analysis}, and, like the $15^\circ$ Bell fault, it is caught by the decision-tree baseline (Section~\ref{sec: baselines}). Every bug whose ideal divergence exceeds the noise floor was either caught or flagged for review, and the extra-Z fault that is invisible on a Bell pair measured in the Z basis was vetoed at $D_B^{\mathrm{log}} = 1.136$ when the same pair was measured in the X basis.

The development suite tells the same story with one difference: its single error is a correct five-qubit Clifford circuit vetoed at $D_B^{\mathrm{log}} = 0.246$, recorded while the backend reported a \texttt{maintenance} status. The identical circuit measured $0.052$ the following day under normal operation and was attributed correctly. We report the error rather than exclude it; it is the only false veto in 95 correct hardware circuits across both devices, and it coincided with a flagged device state.

Two qualifications apply. The held-out families are the ones the training generator was extended with after the seed-0 failure, so the test demonstrates generalization to unseen \emph{instances} on a second device, not to unseen algorithm families. And $n = 30$ gives wide intervals; the point estimates should be read with their brackets. Within those limits, the result is that the model of Section~\ref{sec: training}, with thresholds calibrated on one device's shallow suite, transfers to deeper circuits on a second device with no retuning.

\section{Discussion}
The hardware validation results demonstrate both the promise and the boundaries of physics-informed machine learning for quantum error attribution. 

\subsection{Validation of Physics-Informed Feature Engineering}
\begin{figure}[h]
    \centering
    \includegraphics[width=0.7\linewidth]{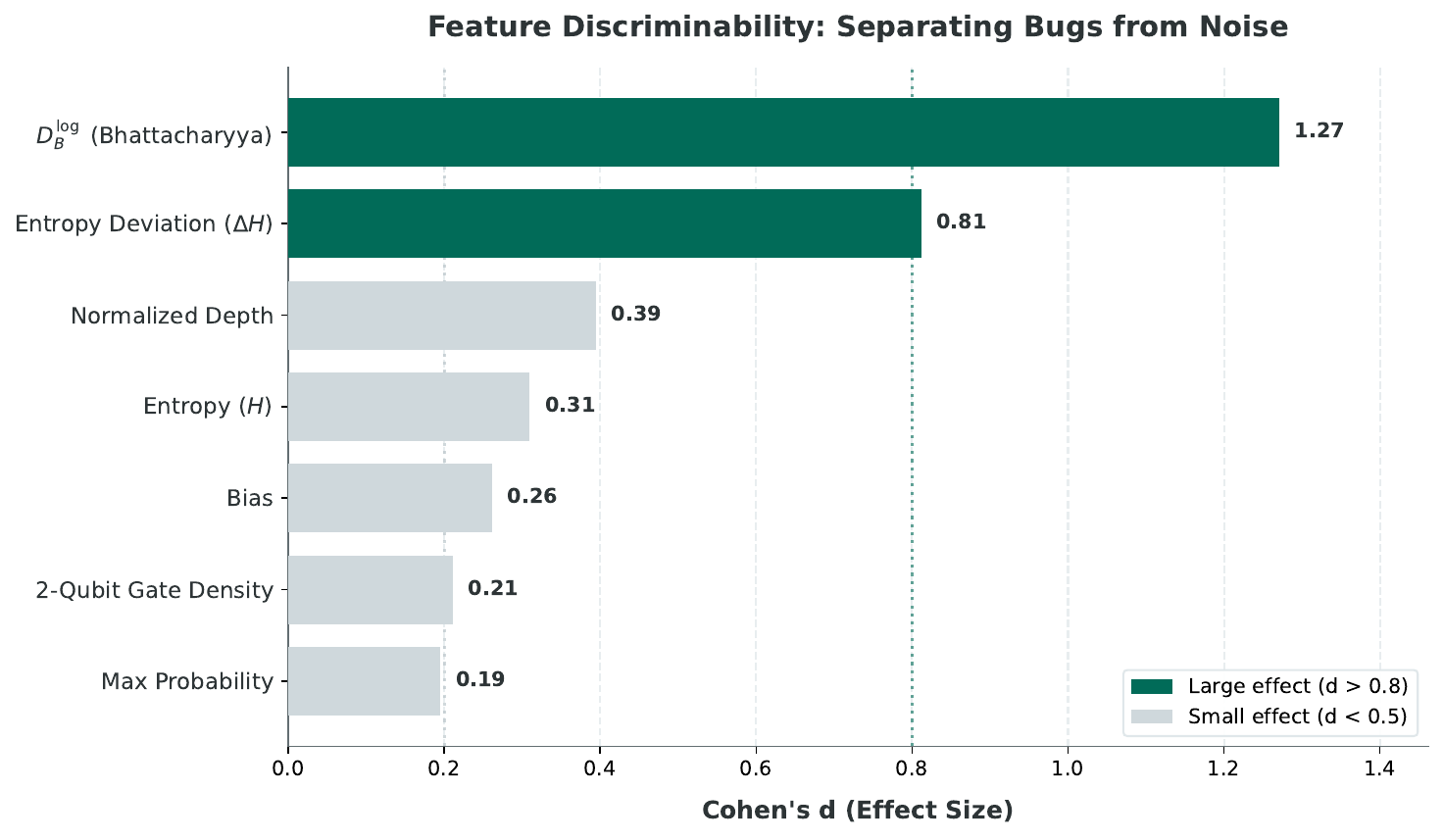}
    \caption{Single-feature discriminability on the 105-circuit primary suite,
    measured as Cohen's $d$ between correct and buggy circuits. The
    log-transformed Bhattacharyya distance and entropy deviation are the only
    features with large effect sizes; the remaining five are individually weak,
    which is why the classifier combines them non-linearly rather than
    thresholding any one of them.}
    \label{fig:discriminability}
\end{figure}

The primary-suite results support the central hypothesis: thermodynamic and topological metrics are robust discriminators for quantum error attribution. As shown in Figure \ref{fig:discriminability}, the log-transformed Bhattacharyya distance exhibits the largest effect size ($d=1.27$), confirming its role as the primary ``topological'' discriminator for logic errors that produce distributions disjoint from the ideal. By Cohen's conventions~\cite{cohenStatisticalPowerAnalysis2013}, this constitutes a large effect. Entropy deviation ranks second ($d=0.81$), supporting the premise that unitary bugs and stochastic noise leave distinct entropic footprints. Cohen's $d$ and the AUC ordering of Table~\ref{tab:feature_auc} agree on the first three positions, with normalized depth, computed on the transpiled circuit, third ($d = 0.39$). The Bhattacharyya distance is also the strongest single discriminator by AUC ($0.858$); the two measures differ only in how they order the weak features.
\begin{table}[h]
\centering
\caption{Single-Feature Discriminative Power on the 105-Circuit Primary Suite (ISA structural features)}
\label{tab:feature_auc}
\small
\begin{tabular}{lrr}
\toprule
\textbf{Feature} & \textbf{AUC} & \textbf{Cohen's $d$} \\
\midrule
Log-transformed Bhattacharyya distance & 0.858 & 1.27 \\
Entropy deviation                      & 0.647 & 0.81 \\
Normalized circuit depth               & 0.625 & 0.39 \\
Bias                                   & 0.586 & 0.26 \\
Shannon entropy                        & 0.580 & 0.31 \\
Maximum probability                    & 0.564 & 0.19 \\
Two-qubit gate density                 & 0.540 & 0.21 \\
\bottomrule
\end{tabular}
\end{table}

These features prove their value in resolving the ``Grover boundary problem'' 
(Section~\ref{sec:grover}). As shown in Figure~\ref{fig:grover_boundary}, 
measuring deviation from \textit{expected} entropy---rather than absolute 
entropy---successfully separates algorithmic intent from error-induced collapse.
The Bhattacharyya veto fired on 27 primary-suite circuits, all of them bugs, and agreed with the classifier on all but one; its value is as a verified hard constraint rather than as a source of accuracy (Section~\ref{sec: veto}).

The remaining four features---bias, maximum probability, raw entropy, and two-qubit gate density---exhibit small individual effect sizes ($d \le 0.31$). This disparity motivates a non-linear classifier; Section~\ref{sec: baselines} quantifies the margin over a linear one, which is modest. While weak in isolation, these features have high \textit{conditional} discriminative power: high bias is evidence of correct operation when entropy deviation is low and of a fault when it is high, though it is not decisive on its own, as the Grover overrotation fault of Section~\ref{sec:grover} shows. Linear models cannot represent such conditional dependencies~\cite{minskyPerceptronsIntroductionComputational}; the fuzzy rule base of ANFIS \cite{jangANFISAdaptivenetworkbasedFuzzy1993} can.

\subsection{The Z-Basis Blindness Limitation}\label{sec: z-basis}

The forensic analysis reveals a fundamental limitation: because the framework relies only on computational-basis measurements, it cannot see phase-only errors. The nine Z-basis-invisible faults of Section~\ref{sec: error_analysis} come from two mechanisms: faults that change only relative phases, and removed gates that act trivially on the tested input state. The second is a limit of test-input coverage and is addressed by varying the input. This section concerns the first.

\subsubsection{The Physics of Phase Invisibility} 
\begin{figure}[h]
    \centering
    \includegraphics[width=0.9\linewidth]{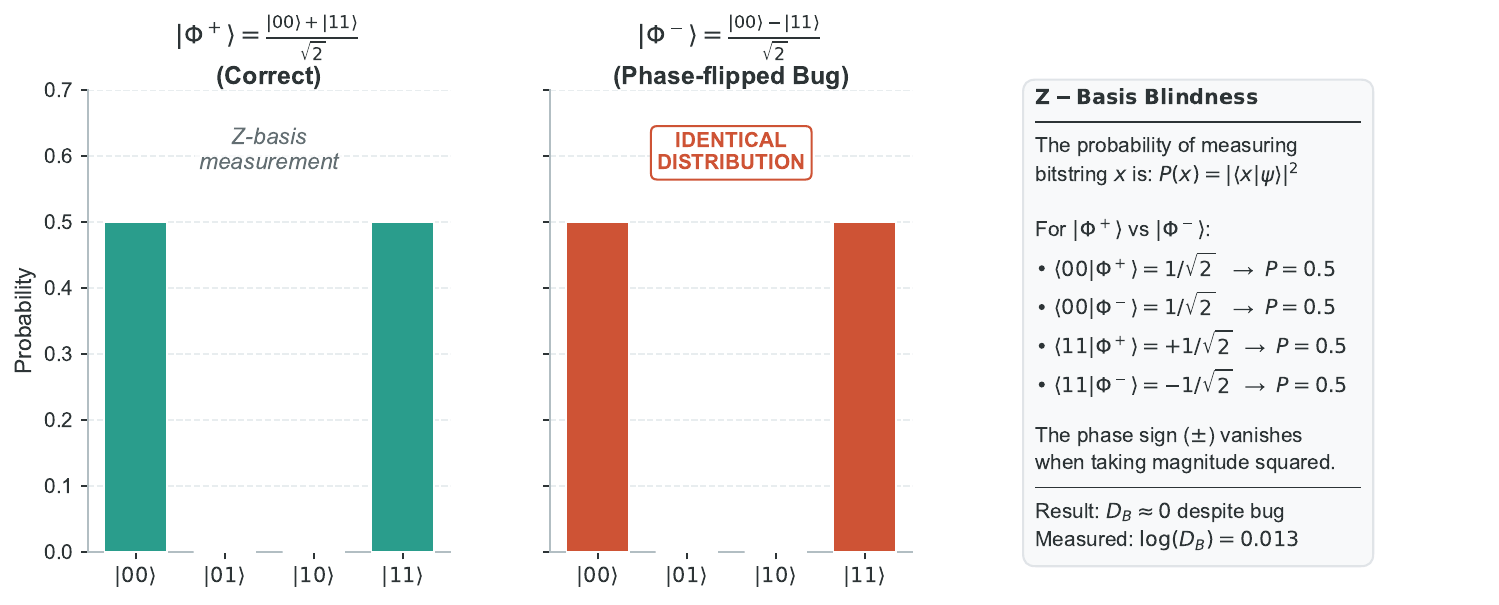}
    \caption{The Z-basis blind spot illustrated with Bell states. Both $|\Phi^+\rangle$ and $|\Phi^-\rangle$ produce identical probability distributions $P(00) = P(11) = 0.5$ when measured in the computational basis; phase information is lost.}
    \label{fig:zbasis_blindness}
\end{figure}

Quantum measurement collapses the state $|\psi\rangle$ onto eigenvectors of the measurement operator. In the Z-basis, we observe $|\langle z|\psi\rangle|^2$ rather than the full amplitude $\langle z|\psi\rangle$. Consider two Bell states:
\begin{align}
    |\Phi^+\rangle &= \frac{|00\rangle+|11\rangle}{\sqrt{2}} \quad \mathrm{(correct)} \\
    |\Phi^-\rangle &= \frac{|00\rangle-|11\rangle}{\sqrt{2}} \quad \mathrm{(phase\text{-}flipped)}
\end{align}
Figure~\ref{fig:zbasis_blindness} illustrates this limitation. Both states produce identical probability distributions when measured in the computational basis: $P(00) = P(11) = 0.5$. The phase information---encoded in the $\pm$ sign---vanishes when we take the modulus squared. X-basis measurement would distinguish them, but this requires additional gate overhead.

\subsubsection{The Role of $T_2$ Dephasing} 

Dephasing does not cause the blind spot. A phase error is invisible in Z-basis counts even on a noiseless device. Dephasing does, however, limit the remedy. The \texttt{ibm\_fez} medians give $T_2 = 102.7\,\mu\text{s}$, against a relaxation-limited ceiling of $2T_1 = 290.4\,\mu\text{s}$, so coherence reaches $35\%$ of the value attainable under $T_1$ alone. Inverting $1/T_2 = 1/(2T_1) + 1/T_\phi$ gives $T_\phi \approx 159\,\mu\text{s}$, so pure dephasing supplies roughly $65\%$ of the transverse decoherence rate. A measurement in a rotated basis recovers a phase error, but with its contrast reduced by the dephasing accumulated over the circuit. Even the deepest circuits studied here run for a small fraction of $T_\phi$, so this reduction is minor, which is consistent with the X-basis Bell result below. The reduction grows with circuit duration, so small phase errors in long circuits would remain hard to resolve even after a basis change.

\subsubsection{Multi-Basis Measurement: An Empirical Test} \label{multibasis}

The Z-basis limitation suggests extending the framework to measurements in more than one basis. We tested this with a 21-feature variant that adds X- and Y-basis Bhattacharyya distances and cross-basis entropy correlations. The variant reached 89.25\% accuracy on its synthetic validation set. On \texttt{ibm\_fez}, however, it misclassified roughly twice as many circuits as the single-basis model trained alongside it. This experiment used the shallow pipeline of Section~\ref{sec: corrections}. Both models were trained on it, and we cannot separate that pipeline's faults from any cost that is specific to multi-basis features, so the result is not evidence against multi-basis diagnostics. The basis rotations add only one layer of single-qubit gates, with a median error of $2.7 \times 10^{-4}$ per gate. The held-out suite shows that a basis change can recover a fault that is otherwise invisible: a Bell pair with an extra Z gate, identical to the correct pair in the Z basis (Section~\ref{sec: error_analysis}), produced $D_B^{\mathrm{log}} = 1.136$ in the X basis and was vetoed. This suggests choosing the measurement basis for the fault class under test, rather than measuring in every basis. Repeating the multi-basis experiment under the training distribution of Section~\ref{sec: training} is left to future work.

\subsection{The Resolution Limit}\label{sec: resolution}
 
Two distinct limits bound what this framework can resolve, and the
\texttt{bell\_angle} cases separate them.
 
The first is sampling. With $N = 4096$ shots, the standard error on a measured outcome probability is $\sigma_p = \sqrt{p(1-p)/N} \leq 0.78\%$, so a fault that shifts an outcome probability by roughly 2.3 percentage points reaches $3\sigma$ separation. This limit is permissive.

The second limit is the sensitivity of the divergence used to score the shift, and it binds much earlier. Suppose probability is redistributed among outcomes that the ideal distribution already occupies, so that $p_i = q_i + \delta_i$ with $\sum_i \delta_i = 0$ and $q_i > 0$. Expanding the coefficient to second order gives
\begin{equation}
    D_B \;\approx\; \frac{1}{8}\sum_i \frac{\delta_i^2}{q_i},
\end{equation}
which, for a shift $\delta$ between two equiprobable outcomes, reduces to $BC \approx 1 - \delta^2/2$. The divergence is quadratic in the deviation, so small shifts are pushed toward the noise floor far more strongly than the sampling statistics alone would suggest. Probability moved outside the ideal support enters linearly instead, and this is what the veto relies on (Section~\ref{sec: veto}).

The two \texttt{bell\_angle} faults illustrate the gap. Their probability shifts, $\delta = \sin(\epsilon)/2$, are $4.36$ and $12.94$ percentage points. Against the sampling floor these are $5.6\sigma$ and $16.6\sigma$, and easy to resolve. Yet their measured $D_B^{\mathrm{log}}$ values are $0.0139$ and $0.0202$, close to the sum of each fault's own contribution ($\approx 0.001$ and $0.009$) and the Bell circuit's noise baseline of $0.0114$. The measured noise floor across the 61 correctly-executed circuits is $0.042$. Both values sit below it, and the classifier attributes both faults to hardware noise with high confidence ($P = 0.97$ and $P = 0.92$).

Solving the exact expression for the fault's own contribution at $D_B^{\mathrm{log}} = 0.042$ gives the detectability condition for this circuit family: $\delta > 0.275$, which corresponds to $\epsilon > 33^\circ$ (near $28^\circ$ once the circuit's baseline is included). The two faults therefore produce shifts that are large relative to sampling noise but small relative to the divergence that hardware noise itself produces on correctly-executed circuits. Unlike the phase invisibility of Section~\ref{sec: z-basis}, which is a property of computational-basis measurement, this bound is empirical. It reflects where this device's noise floor sits, and it would move on a lower-noise processor. It would not move with a larger shot budget. The expected sampling contribution to $D_B$ is approximately $(K-1)/(8N)$ for $K$ occupied outcomes. At 4{,}096 shots this is $3 \times 10^{-5}$ for a Bell pair and below $10^{-3}$ even for a uniform five-qubit output, one to three orders of magnitude below the floor.

\begin{figure}
    \centering
    \includegraphics[width=0.80\linewidth]{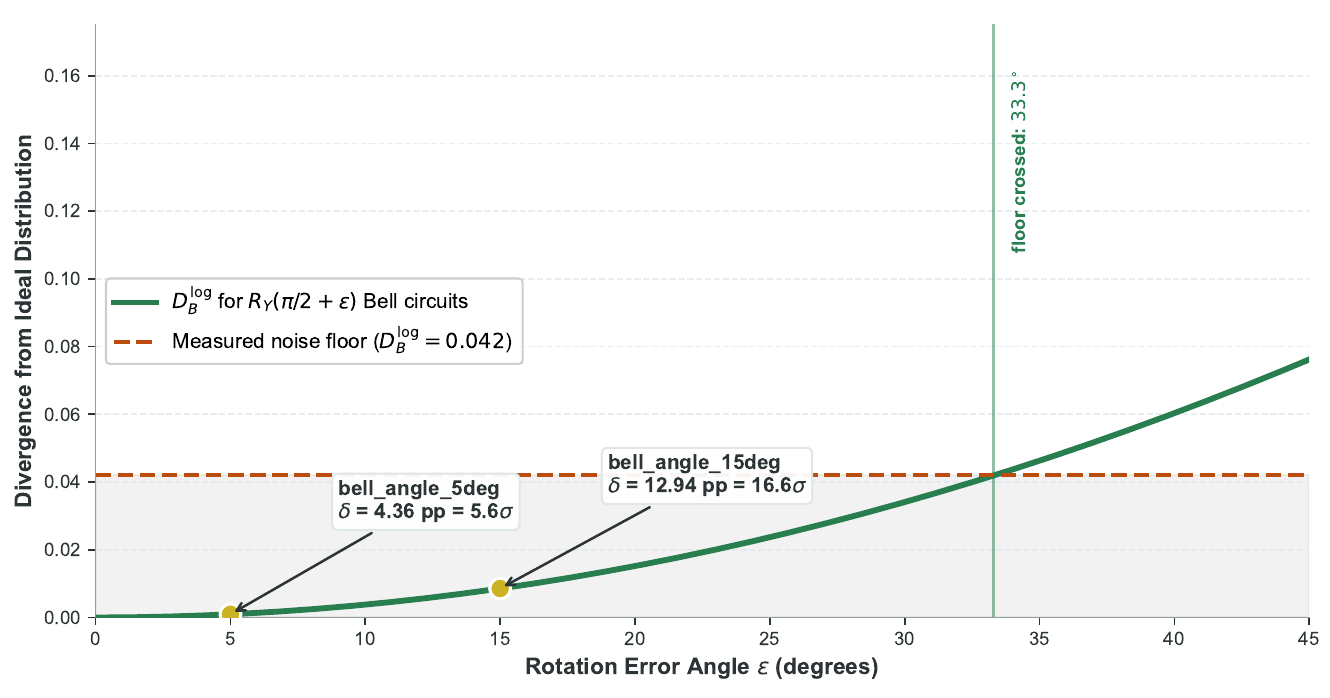}
    \caption{Resolution limit imposed by the Bhattacharyya feature. For
$R_Y(\pi/2+\epsilon)$ Bell circuits, $D_B$ grows quadratically in the
probability shift, so the divergence stays below the measured noise floor
($0.042$, shaded) until $\epsilon \approx 33^\circ$. Plotted values are each
fault's own contribution; the measured divergences, $0.0139$ and $0.0202$,
also include the Bell circuit's noise baseline.}
    \label{fig:resolution_limit}
\end{figure}

Figure~\ref{fig:resolution_limit} plots the relationship: the two faults sit
well above the sampling floor but well below the divergence floor.

The floor is a limit of the Bhattacharyya feature on its own, and entropy deviation adds information just above it. Across the three hardware suites, 36 correct circuits have $D_B^{\mathrm{log}}$ between $0.014$ and $0.060$, and every one of them has $\Delta H$ above $0.28$. When noise moves an output this far from the ideal, it also spreads it. The two below-floor faults that ANFIS misses and the decision-tree baseline catches, the $15^\circ$ Bell fault and the ansatz fault of the held-out test, have $\Delta H$ of $0.107$ and $0.074$. Their outputs moved without spreading, which is the purity-preserving signature of a coherent error (Section~\ref{sec: feature 6}). The training set shows the same one-sided pattern: its 393 noise samples in this range all have $\Delta H$ of at least $0.18$, while its bug samples in the range fall on both sides of that value. Low entropy deviation above a circuit's clean divergence is therefore evidence of a fault, but high entropy deviation is not evidence against one. The pattern is physically motivated and holds on both devices, but it rests on few bugs, so a rule or feature that targets it would need to be tested on a new held-out suite.

A potential mitigation is incorporating temporal features: hardware noise
fluctuates over calibration cycles while software bugs are deterministic
\cite{burnettDecoherenceBenchmarkingSuperconducting2019}. Measuring
distribution stability across time could disambiguate small coherent bugs from
calibration jitter.

\subsection{Practical Deployment}

In practical deployment, the framework functions as a pre-mitigation gatekeeper within quantum software development pipelines. Consider a continuous integration workflow: a developer commits circuit code, which is transpiled and executed on hardware. The framework then analyzes the output distribution and routes the result through one of three paths.

Circuits classified as \texttt{HARDWARE\_NOISE} proceed to error mitigation (ZNE, PEC, or dynamical decoupling), with the build conditionally passing and a warning logged. This path shows only that no fault was detected in the measured basis; it does not certify the circuit. Eleven of the 44 primary-suite bugs, including all nine Z-basis-invisible faults, would pass it (Section~\ref{sec: error_analysis}). 
Circuits classified as \texttt{SOFTWARE\_BUG} trigger an immediate build failure, and the fuzzy rule explanation is returned to the developer. For example, for a circuit vetoed at $D_B^{\mathrm{log}} = 0.30$ on which rule R9 fires most strongly, the explanation reads: high entropy deviation and high divergence on a shallow circuit, the signature of a coherent gate substitution rather than accumulated noise.

The manual-review overhead is therefore small: at most one circuit in thirty on the suites reported here, and none on the primary suite. The value of the abstention mechanism lies in the cases it does flag, which were in every instance bugs whose divergence sat near the device's noise floor.

\subsection{Interpretability as a Debugging Tool}

In quantum debugging, interpretability is not a secondary benefit but a functional requirement. The utility of the framework lies in exposing the attribution rationale---developers receiving a \texttt{SOFTWARE\_BUG} flag need to understand \textit{why} to guide debugging efforts. Table~\ref{tab:extracted_rules} demonstrates this concretely: the top five rules extracted from the trained model map directly onto distinct physical fault categories.

The ANFIS architecture's ability to extract human-readable rules turns each
attribution into a diagnostic pointer:
\begin{itemize}
    \item \textbf{R9 --- high $\Delta H$ and high $D_B^{\log}$ on a shallow,
    lightly entangled circuit:} a coherent gate substitution. The limited
    noise budget cannot account for the discrepancy, so the search should
    start at the gates, not the device.
    \item \textbf{R12 --- low entropy, high bias, high $D_B^{\log}$:}
    probability concentrated on the wrong basis state. Look for a unitary
    error that collapses the intended superposition.
    \item \textbf{R6 --- low entropy and high bias, but low $\Delta H$ and low
    $D_B^{\log}$:} peaked by design rather than by accident. No bug; the
    algorithm intended a concentrated output.
\end{itemize}
The contrast between R12 and R6 is the diagnostically useful one: both fire on
peaked, high-bias outputs, and they are separated by entropy deviation and
divergence alone. A developer receiving a \texttt{SOFTWARE\_BUG} flag from R12
knows the output is peaked at the wrong state; one whose circuit fires R6
knows the peak is intended.

\subsection{Relation to Adjacent Systems}
Three systems operate on quantum output data but answer different
questions; the distinctions clarify where attribution sits.
\begin{itemize}
    \item \textbf{QOIN \cite{muqeetMitigatingNoiseQuantum2024c}:} QOIN achieves $\sim 80\%$ noise reduction through learned filtering; 
    it focuses on cleaning output distributions rather than diagnosing 
    error sources. Our framework produces explicit classifications enabling 
    different remediation pathways. The approaches are complementary---QOIN 
    could preprocess distributions before our attribution step.
    
    \item \textbf{AlphaQubit \cite{bauschLearningHighaccuracyError2024b}:} 
    AlphaQubit excels at neural error correction within QEC codes, but operates on the premise that the logical circuit is correct. Our framework operates at a higher abstraction layer, performing diagnostic triage to catch software bugs that a decoder would otherwise mistakenly preserve as valid logical states.
    
    \item \textbf{Statistical assertions \cite{huangStatisticalAssertionsValidating2019a}:} 
    Statistical assertion frameworks require developers to manually specify expected properties (e.g., ``output should be a superposition''). Our approach needs no developer-specified properties and learns discriminative signatures from training data. It does, however, require the ideal output distribution as a reference, which assertions do not. It trades the specificity of assertions, and their independence from a reference, for automated use across circuit structures.
\end{itemize}

\subsection{Baseline Model Comparison}\label{sec: baselines}

To assess whether the neuro-fuzzy architecture is necessary, we compared ANFIS
against logistic regression, an RBF-kernel support vector machine, and a
decision tree. All four models were trained on the same 1,983-sample
synthetic dataset of Section~\ref{sec: training}, and all four were evaluated
on the same 105 hardware circuits under identical two-stage decision logic:
the Bhattacharyya veto followed by the $0.35$/$0.70$ abstention bands.

Effective accuracy alone is not a sufficient basis for this comparison,
because it rewards abstention: a classifier that abstains on every input
attains $100\%$ effective accuracy and $0\%$ strict accuracy. We therefore
report the comparison in the risk--coverage terms standard in classification
with rejection~\cite{chowOptimumRecognitionError1970}: \emph{coverage} is
the fraction of circuits on which the model commits to a decision, and
\emph{selective accuracy} is the accuracy on that committed subset.

\begin{table}[h]
\centering
\caption{Baseline comparison on the 105-circuit primary suite. All four models share one training set and one decision rule.}
\label{tab:baselines}
\small
\begin{tabular}{lrrrr}
\toprule
\textbf{Model} & \textbf{Coverage} & \textbf{Selective acc.} & \textbf{Strict acc.} & \textbf{Distinct $p$} \\
\midrule
Logistic regression & 98.1\% & 88.3\% & 86.7\% & 86 \\
SVM (RBF kernel)    & 97.1\% & 88.2\% & 85.7\% & 78 \\
Decision tree       & 100.0\% & 90.5\% & 90.5\% & 2 \\
ANFIS (this work)   & 100.0\% & 89.5\% & 89.5\% & 92 \\
\bottomrule
\end{tabular}
\end{table}

Table~\ref{tab:baselines} shows that, with the training data of Section~\ref{sec: training}, the four models are within five percentage points of one another, and all four commit on at least 97\% of circuits. The nine Z-basis-invisible bugs of Section~\ref{sec: error_analysis} bound every model at 96 correct: the decision tree reaches 95, ANFIS 94, and logistic regression 91. On training data that still carried the faults of Section~\ref{sec: corrections}, the same comparison put ANFIS ahead by 7 to 23 points of strict accuracy. That margin came from the faulty data, not from the task, and it disappears once the three requirements of Section~\ref{sec: corrections} are met.

The tree's ten errors are a subset of the eleven ANFIS errors, and it attributes the $15^\circ$ Bell fault correctly through the entropy-deviation pattern of Section~\ref{sec: resolution}. We therefore scored the baselines on both held-out suites as well, with the same frozen thresholds (Table~\ref{tab:baselines_holdout}). Over all 165 hardware circuits, the decision tree is correct on 152 and ANFIS on 150, and both make 13 errors; the difference is two circuits and is not statistically significant (exact McNemar test, $p = 0.5$). The models differ in how they fail rather than in how often. The Trotter angle error, whose measured divergence sits just above the noise floor, appears in both held-out suites. ANFIS abstains on it both times ($P(\mathrm{noise}) = 0.59$ and $0.65$). The decision tree attributes it to noise both times with $P(\mathrm{noise}) = 1$, a confident error that nothing in its output can flag.
\begin{table}[h]
\centering
\caption{Strict-correct counts for all four models on the three hardware suites, under the same veto and thresholds. The held-out suites have 30 circuits each.}
\label{tab:baselines_holdout}
\small
\begin{tabular}{lrrrrrr}
\toprule
\textbf{Model} & \texttt{ibm\_fez} & \textbf{Seed 0 (dev.)} & \textbf{Seed 2 (test)} & \textbf{All 165} & \textbf{Abstain} & \textbf{Errors} \\
\midrule
Logistic regression & 91 & 26 & 26 & 143 & 5 & 17 \\
SVM (RBF kernel)    & 90 & 29 & 27 & 146 & 5 & 14 \\
Decision tree       & 95 & 28 & 29 & 152 & 0 & 13 \\
ANFIS (this work)   & 94 & 28 & 28 & 150 & 2 & 13 \\
\bottomrule
\end{tabular}
\end{table}
What the comparison does establish is structural. The decision tree produces two distinct output probabilities across all 105 circuits, so no input can fall inside the abstention band and the model cannot express graded confidence at any threshold; this is intrinsic to axis-parallel partitioning with leaf-frequency estimates. ANFIS produces 92 distinct values and uses the band where the held-out suites show it should, on bugs near the noise floor. Logistic regression and the SVM also produce graded outputs, but they express the boundary as a single linear weighting or a kernel expansion, and the tree's splits can be read while its outputs cannot be graded.

These results do not support a claim that ANFIS is uniquely accurate. With training data free of those faults, simple models are competitive and most of the attribution signal is carried by two features. They support a narrower claim: among the models considered, ANFIS is the one that combines competitive accuracy with graded confidence and a rule base that can be read as physical conditions.

\section{Conclusion}
This paper addressed the attribution problem in quantum software engineering: determining whether unexpected program outputs arise from software bugs requiring debugging or hardware noise amenable to mitigation. We presented a physics-informed neuro-fuzzy framework that combines learned classification with a hard, empirically calibrated constraint. The Bhattacharyya veto rejects noise attributions when a measured distribution diverges from the ideal by more than the device's noise produces under normal operation. The entropy deviation feature resolves a key ambiguity---distinguishing algorithms that produce peaked distributions by design from bugs that collapse superpositions by accident---and the trained rule base encodes that resolution, the veto's own logic, and the behavior of deep entangled circuits under noise as explicit, readable rules. The framework attained 89.5\% accuracy on the primary suite and 93.3\% strict accuracy on a held-out suite scored once with every threshold frozen. Every residual error on the primary suite is a bug whose ideal computational-basis distribution equals the intended circuit's (nine of eleven) or whose shift lies below the device's divergence floor (two); the failures observed with a shallower training pipeline were traced to specific properties of that pipeline (Section~\ref{sec: corrections}). Two boundaries constrain single-basis diagnostics. Phase-only errors, and faults that act trivially on the tested input, are undetectable from computational-basis counts by any method; a basis chosen for the fault class can recover them, as the X-basis Bell measurement of the held-out suite shows. Rotation errors whose probability shift is large relative to sampling noise but small relative to the divergence that noise itself induces---below $\delta \approx 0.28$ for Bell circuits on this device---are compressed beneath the floor of the Bhattacharyya feature. 

\subsection*{Future Directions}

\paragraph{Debugging Through the QEC Abstraction Barrier:} 
As quantum computing transitions toward fault tolerance, the bug-versus-noise 
distinction changes fundamentally. When error correction is active, physical 
errors are masked by decoders, and any residual logical error could be a 
software bug, a decoder failure, or noise exceeding the correction threshold. 
Google's recent below-threshold demonstration identified rare correlated error events that limited logical performance---events 
that current frameworks cannot attribute to specific causes 
\cite{googlequantumaiandcollaboratorsQuantumErrorCorrection2025a}. Extending attribution to logical qubits requires solving the ``abstraction barrier'' problem: syndrome information consumed by decoders hides the raw error data that our physics-informed features rely on.

\paragraph{Cross-Platform Generalization:}
Our validation is limited to superconducting processors with their characteristic $T_1$/$T_2$ decay and ZZ crosstalk. Other platforms exhibit different error mechanisms: neutral atom systems can convert dominant errors into detectable erasures via metastable encoding \cite{wuErasureConversionFaulttolerant2022, reichardtFaulttolerantQuantumComputation2025}, while trapped ions suffer measurement-induced heating orders of magnitude faster than ambient rates \cite{rasmussonMeasurementinducedHeatingTrapped2024}. Recent cross-platform benchmarking of 24 QPUs from 6 vendors revealed that non-local crosstalk and thermal effects vary substantially by architecture \cite{montanez-barreraEvaluatingPerformanceQuantum2025}. The question is not whether our features transfer directly---they likely do not---but whether the underlying principle (thermodynamic signatures distinguish unitary bugs from 
stochastic noise) holds across modalities.

\begingroup
\footnotesize 
\bibliographystyle{unsrturl}
\bibliography{bibfile}
\endgroup

\end{document}